\documentclass[review,number,sort&compress,3p,times]{elsarticle}

\usepackage{hyperref}

\usepackage{lineno}
\usepackage{color}

\begin{document}

\begin{frontmatter}

\title{ The high voltage system with pressure and temperature corrections for the novel MPGD-based photon detectors of COMPASS RICH-1}

\author[mymainaddress,address3]{J.~Agarwala}
\author[mymainaddress]{M.~Bari}
\author[mymainaddress,address2]{F.~Bradamante}
\author[mymainaddress,address2]{A.~Bressan}
\author[mymainaddress,address2]{C.~Chatterjee}
\author[mymainaddress,address3]{A.~Cicuttin}
\author[mymainaddress,address2]{P.~Ciliberti}
\author[mymainaddress,address3]{M.~Crespo}
\author[mymainaddress]{S.~Dalla~Torre}
\author[mymainaddress]{S.~Dasgupta}
\author[mymainaddress]{B.~Gobbo}
\author[mymainaddress]{M.~Gregori}
\author[mymainaddress]{G.~Hamar \fnref{presentaddress}}
\author[mymainaddress]{S.~Levorato}
\author[mymainaddress,address2]{A.~Martin}
\author[mymainaddress]{G.~Menon}
\author[mymainaddress,address2]{L.B.~Rizzuto\fnref{presentaddress2}}
\author[mymainaddress,address3]{Triloki}
\author[mymainaddress]{F.~Tessarotto}
\author[mymainaddress]{Y.~X.~Zhao \corref{mycorrespondingauthor}}
\cortext[mycorrespondingauthor]{Corresponding author}
\ead{yuxiang.zhao@ts.infn.it}


\address[mymainaddress]{INFN Sezione di Trieste, Padriciano 99, Trieste, Italy}
\address[address2]{University of Trieste, Trieste, Italy}
\address[address3]{Abdus Salam ICTP, Trieste,Italy}
\fntext[presentaddress]{Present address: Wigner Research Center for Physics, Budapest, Hungary}
\fntext[presentaddress2]{Present address: Josef Stefan Institute, Ljubljana, Slovenia}

\begin{abstract}

The novel MPGD-based photon detectors of COMPASS RICH-1 consist of large-size hybrid 
MPGDs with multi-layer architecture including two layers of Thick-GEMs and a bulk 
resistive MicroMegas. The top surface of the first THGEM is coated with a CsI film 
which also acts as photo-cathode. These detectors have been successfully in 
operation at COMPASS since 2016. Concerning bias-voltage supply, the Thick-GEMs are 
segmented in order to reduce the energy released in case of occasional discharges, 
while the MicroMegas anode is segmented into pads individually biased with positive 
voltage while the micromesh is grounded. In total, there are about ten different electrode 
types and more than 20000 electrodes supplied by more than 100 HV channels, where 
appropriate correlations among the applied voltages are required for the correct 
operation of the detectors. Therefore, a robust control system is mandatory, 
implemented by a custom designed software package, while commercial power supply 
units are used. This sophisticated control system allows to protect the detectors 
against errors by the operator, to monitor and log voltages and currents at 1 Hz 
rate, and automatically react to detector misbehaviour. In addition, a voltage compensation
system has been developed to automatically adjust the biasing voltage according to 
environmental pressure and temperature variations, to achieve constant gain over time.
This development answers to a more general need. In fact, voltage compensation is 
always a requirement for the stability of gaseous detectors and its need is 
enhanced in multi-layer ones. 

In this paper, the HV system and its performance are described in details, as 
well as the stability of the novel MPGD-based photon detectors during the physics 
data taking at COMPASS.
\end{abstract}


\end{frontmatter}


\section{Introduction}
\label{sec:introduction}

The RICH-1 detector~\cite{rich-1_1,rich-1_2,rich-1_3}
of the COMPASS experiment~\cite{compass_1,compass_2} is a Ring Imaging Cherenkov 
detector with a 3~m long gaseous C$_4$F$_{10}$ radiator, 21~m$^2$ wide focusing 
VUV mirror surface and Photon Detectors (PD) covering a total active area of 
5.5~m$^2$. In the 2015-2016 winter shut-down of the COMPASS experiment, 
around 1.4 m$^2$ of the active area of the photon detectors was upgraded by novel 
detectors based on MPGD technologies \cite{hybrids_at_compass}. 
Four new PDs (unit size: 600$\times$600~mm$^2$) replace the previously used
MWPC-based PDs in order to cope with the challenging efficiency and stability requirements 
of the new COMPASS measurements. In fact, COMPASS goal is to deal with
trigger rates up to O(10$^5$)~Hz and beam rates up to O(10$^8$)~Hz and the new 
detector architecture guarantees data taking stability.

The high voltage system for the upgraded detectors and its performance,  
as well as its significance on the detector stability are the focus of this paper.

\begin{table*}
\begin{center}
\begin{tabular}{|c|| c |c | c | c | c | c | c | c |}
\hline
Electrode & Protection & Drift & THGEM1 & THGEM1 & THGEM2 & THGEM2 & mesh & MM anode 
\\ & wire plane & wire plane & top & bottom & top & bottom & & \\ \hline  \hline
Voltage & -300 V & -3520 V & -3320 V & -2050 V & -1750 V & -500 V & grounded & +620 V \\ \hline
Number of & & & & & & & & \\
HV channels & 1 & 1 & 4 & 4 & 4 & 4 & 0 & 4 \\
per detector & & & & & & & & \\ \hline
\end{tabular}
\caption{Typical  voltages applied to the electrodes of the hybrid photon detector during 
operation at COMPASS. The required number of HV channels per detector (600x600 mm$^2$) 
are also indicated (details in the text).
}
\label{tab:voltage_number}
\end{center}
\end{table*}

\begin{figure}
\begin{center}
\resizebox{0.5\textwidth}{!}{ \includegraphics{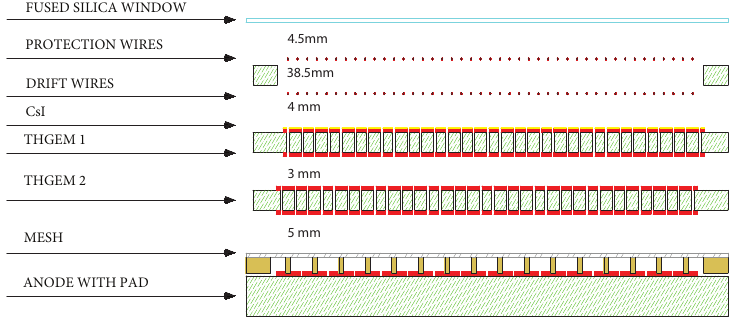} }
\caption{Sketch of the hybrid single photon detector: two staggered 
THGEM layers are coupled to a bulk MicroMegas. The drift wire and 
the protection plane are visible. Distances between the quartz window 
and between the electrodes are indicated too. 
Image not to scale.
}
\label{fig:Hybrid_architecture}
\end{center}
\end{figure}

\begin{figure}
\begin{center}
\resizebox{0.51\textwidth}{!}{ \includegraphics{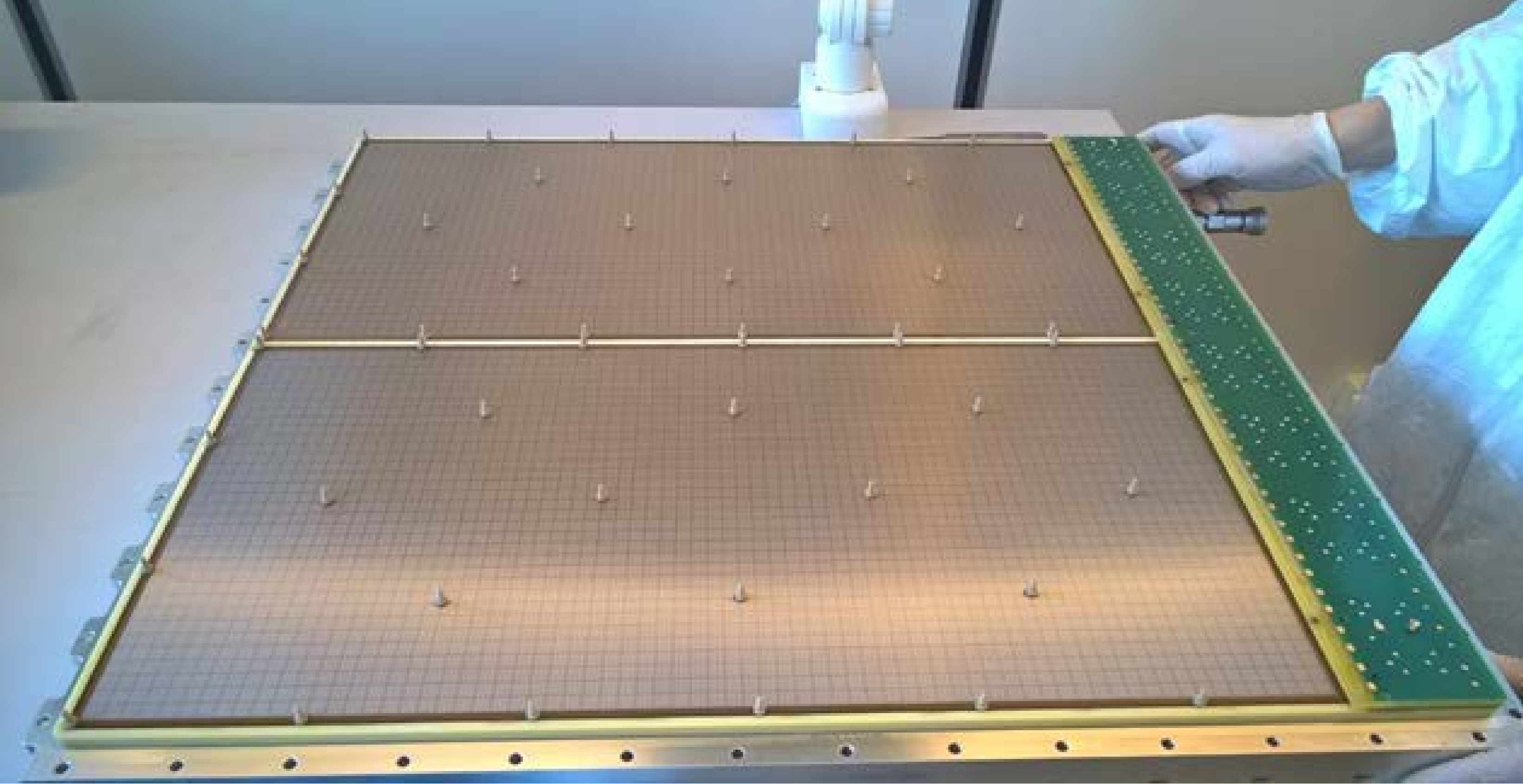} }
\resizebox{0.51\textwidth}{!}{ \includegraphics{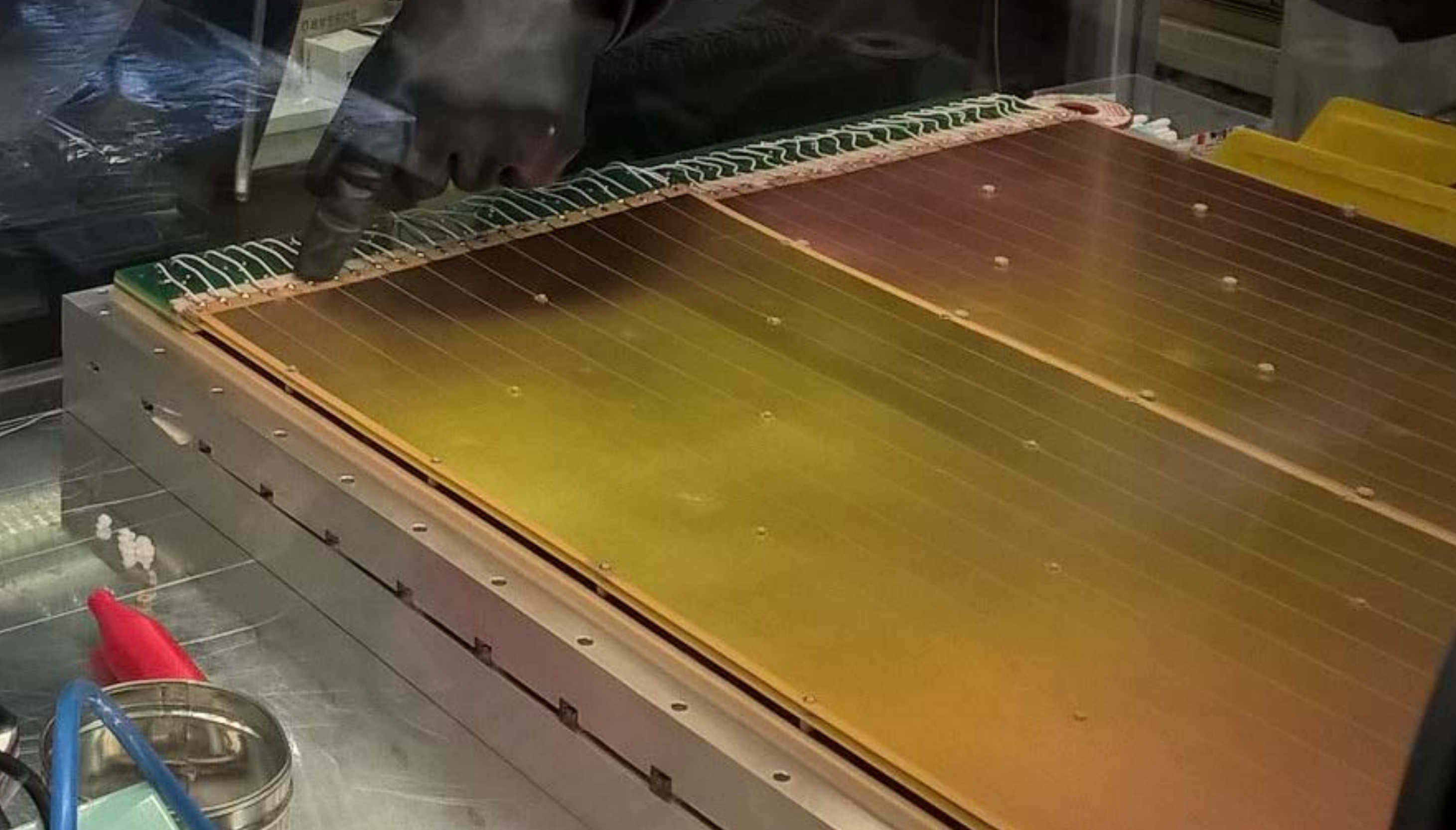} }
\caption{Pictures showing two units (600$\times$300~mm$^2$) of MMs (top picture) 
and THGEMs (bottom picture) in one single detector (600$\times$600~mm$^2$).
}
\label{fig:mm_THGEM_units}
\end{center}
\end{figure}

\begin{figure*}
\begin{center}
\resizebox{0.75\textwidth}{!}{ \includegraphics{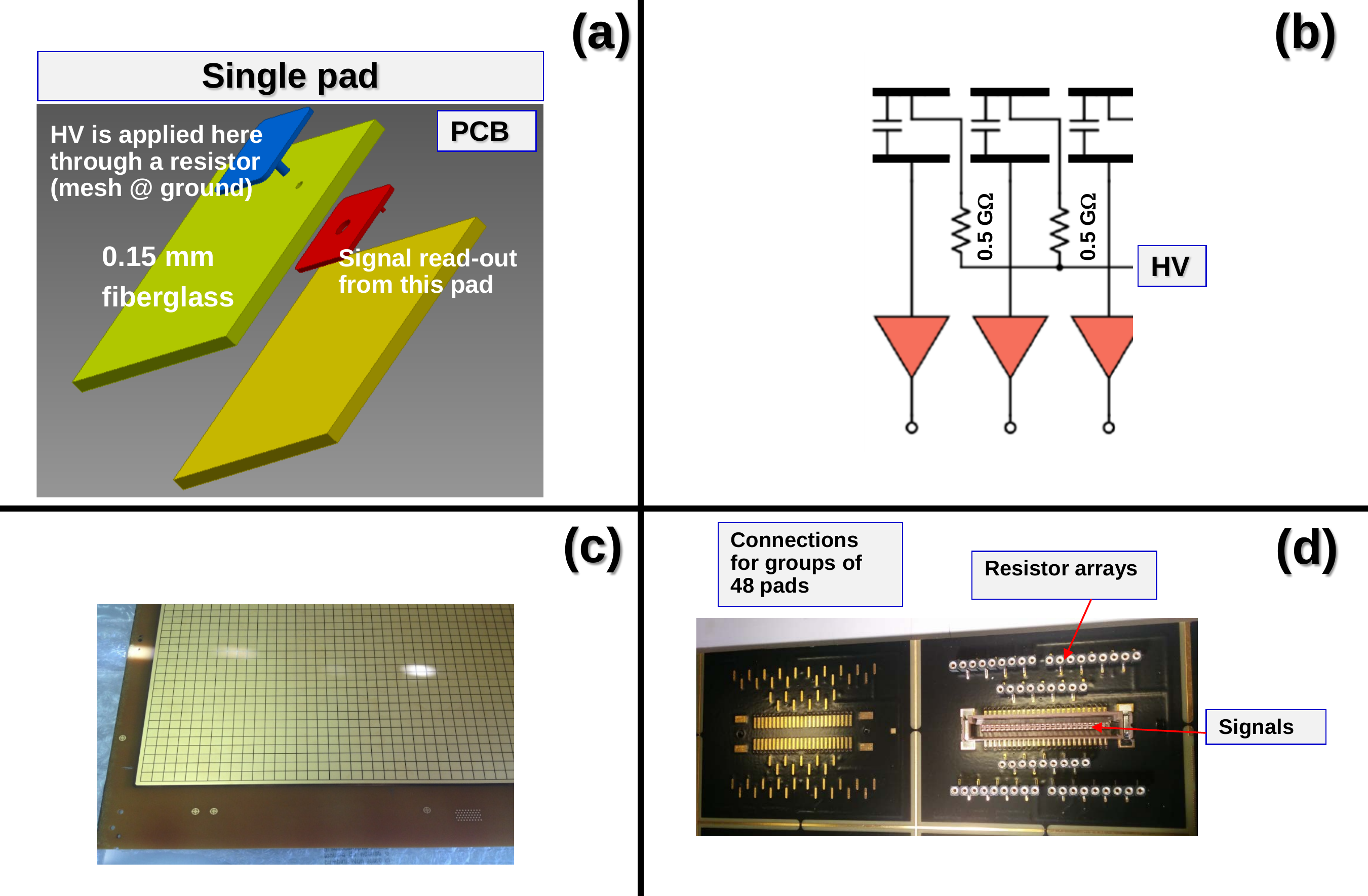} }
\caption{(Color online) The resistive MM by discrete elements.
(a)Sketch of the PCB layers illustrating the principle of readout design. The 
blue (external) pad is the anode electrode of the MM, the red (internal) pad is embedded in the PCB
and the signal is transferred from the blue to the red pad by capacitive coupling.
(b)The principle is illustrated by the electrical scheme. The top elements of the capacitors
are the pad forming the MM anode (blue (external) pad in (a)), the bottom elements of the capacitors
(red (internal) pad in (a)) are connected to the front-end electronics.
(c)Picture of the MM anode PCB, front view.
(d)Picture of the MM anode PCB, rear view. The connectors serving 48 pads are 
grouped together, the signal connectors and resistor array connectors are labeled.
}
\label{fig:resistiveMM}
\end{center}
\end{figure*}

\begin{figure}
\begin{center}
\resizebox{0.4\textwidth}{!}{ \includegraphics{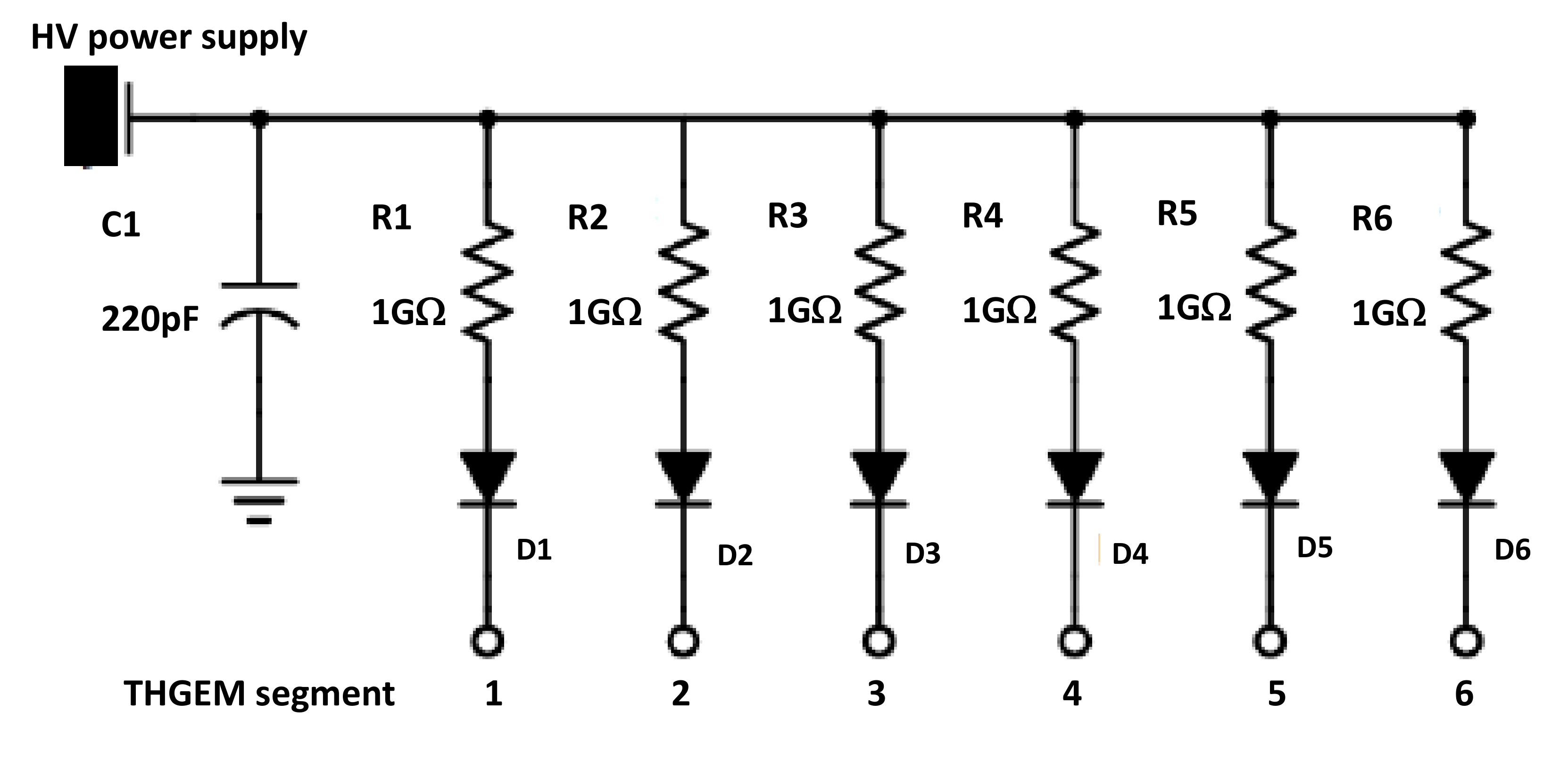} }
\caption{Scheme of the voltage distribution of the top (bottom) face of a THGEM sector
containing 6 segments.
}
\label{fig:THGEM_HV_distributor}
\end{center}
\end{figure}

\section{The architecture of the photon detectors}
\label{sec:architecture} 

The architecture of the novel detectors results from a seven-year R\&D activity~\cite{ourRD_1,ourRD_2,ourRD_3,ourRD_4} 
and it concludes in a hybrid MPGD arrangement (Fig.~\ref{fig:Hybrid_architecture}): two 
layers of THick GEMs (THGEM)~\cite{thgem_1,thgem_2,thgem_3,thgem_4} are coupled to a MicroMegas (MM)~\cite{mm}.
The top surface of the first THGEM is coated with a CsI film which acts as a reflective photo-cathode.
The MM has a pad segmented anode, where the pad size is 7.5$\times$7.5~mm$^2$ with 8~mm pitch. 
Each hybrid 600$\times$600~mm$^2$ detector is built by
two 600$\times$300~mm$^2$ units arranged side by side within a single detector, as shown in 
Fig.~\ref{fig:mm_THGEM_units}, while the protection and drift wire planes are common for the two units.

The MM is built with the bulk technology~\cite{bulk-mm}. The resistive MM concept has been adopted and established
by an original implementation making use of discrete elements: HV is applied to the anode pads,
each one protected by an individual resistor, while the signals are collected from a second set of
pads, parallel to the first ones, embedded in the anode PCB, where the signal is transferred by
capacitive coupling (Fig.~\ref{fig:resistiveMM}). 
Therefore, in the MM stage, the segmentation of the electrodes is very fine: each single pad acts 
as an almost independent electrode, thanks to the large-value resistance (470~M$\Omega$) that 
decouples it from the other pads. The 4760 pads of a detector are supplied by four HV units, 
each one providing bias to 1190 pads.

Both surfaces of each THGEM unit (600x300~mm$^2$) are divided into 12 electrodes, referred 
to as segments in the following. The segments of the top (bottom) THGEM face 
are grouped by six forming two sectors per THGEM unit, namely four per THGEM plane in a detector.
A sector is powered via a single HV channel through a distribution scheme shown in
Fig.~\ref{fig:THGEM_HV_distributor}. The segment electrodes are almost independent thanks 
to the decoupling resistors, while the fast diodes prevent the current flow from a segment to 
the others in case of an occasional discharge, which acts as a local temporary short. 
In addition, the THGEM segmentation offers the opportunity to isolate and power independently 
a few electrically unstable segments.

The detectors are operated with the gas mixture Ar~:~CH$_4$~=~50~:~50, selected for 
optimal extraction of the photo-electrons from the converting CsI film to the gaseous
atmosphere~\cite{ourRD_1,ourRD_2,ourRD_3,ourRD_4}.

The typical voltages applied to the detector electrodes are (with reference to electrodes as 
indicated in Fig.~\ref{fig:Hybrid_architecture}) 1270~V across THGEM1, 1250~V across THGEM2, 
and 620~V to bias the MM (Tab.~\ref{tab:voltage_number}). The drift field above the first THGEM is 500~V/cm, the transfer field 
between the two THGEMs is 1000~V/cm and the field between the second THGEM and the MM micromesh 
is 1000~V/cm. This corresponds to effective gain-values for the three multiplication layers 
around 12, 10 and 120. These values include electron transfer efficiency.

\begin{figure*}
\begin{center}
\resizebox{0.95\textwidth}{!}{ \includegraphics{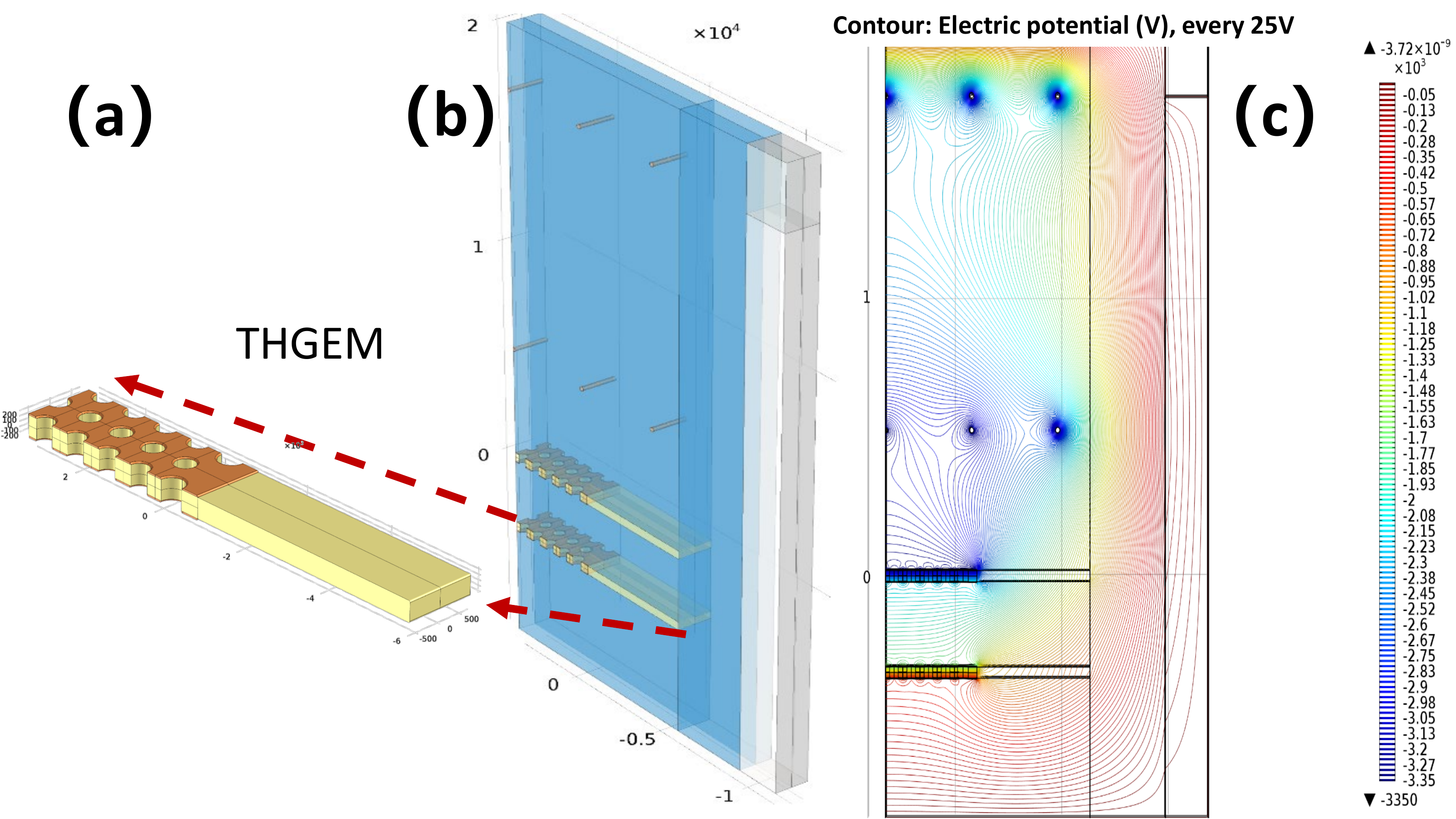} }
\caption{Calculation of the electric field configuration at the lateral edges of the 
detector before introducing the auxiliary electrodes. (a) THGEM element used 
in the calculation. (b) Elements of THGEM1 and THGEM2 and a portion of 
the detector frame. (c) electric potential contour.
}
\label{fig:lateral_electric_field}
\end{center}
\end{figure*}

The detector frames are very near to the active surface in order to reduce the dead area between 
adjacent detectors. These frames are made by aluminum
internally covered with an insulator layer of 
Tufnol 6F/45\footnote{by Tufnol Composites Ltd, 
76 Wellhead Ln, Birmingham B42 2TN, UK.} 
to prevent discharges from the multiplication electrodes 
to the frames. Nevertheless, the presence of the frames, obviously properly grounded, can modify 
the electric field at the detector edges. 
The calculated electric field maps are shown in
Fig.~\ref{fig:lateral_electric_field}; they result in a loss of efficiency 
and non-stability in the lateral portion of the 
detectors. The correct electric field configuration is restored by auxiliary electrodes at the 
lateral sides of the detector frames, embedded below the insulator layers. 
Two electrodes, biased at different appropriate voltages, are located at each detector 
lateral edges for a total number of four electrodes per detector.

\section{Requirements for the high voltage system}
\label{sec:requirements}

The requirements for the HV system originate from specific requests related to MPGDs in 
general and from the architecture of the novel detectors (Sec.~\ref{sec:architecture}).
Current monitoring with fine resolution is needed to control MPGD operation.
The multistage structure of the detectors with three transfer and three 
amplification regions imposes that the main system elements of the HV system are the
relative voltages between the cascaded electrodes and not the individual electrodes
themselves. Therefore, the applied voltages are highly correlated. 
Consequently, the control of the voltages is not trivial, while the electrode
segmentation adds further elements of complexity. These considerations dictate the 
first and more important requirement: the automatized control of a complex system 
in order to prevent errors by the operator.

The complexity of the system also requires to implement automatic protocols to react 
to occasional misbehaviour of the detectors and to log regularly and at relatively high 
frequency the relevant parameters, namely the voltages and currents supplied 
by the HV units.

Changes in pressure and temperature cause gain variations in all gas detectors, which are more
severe in multistage detectors because the variations in the amplification stages are combined
constructively. Therefore, the HV system has to include sensors to monitor the
environmental parameters and protocols to properly modify the applied voltage to
compensate for the effect of the parameter variation.

Last, a user-friendly graphical interface is needed to support the operator 
in controlling the whole system, including about 140 HV power supply channels.
 
\section{The hardware components of the HV system}
\label{sec:hardware}

\subsection{The power supply units}
\label{sec:power-supply}

Commercial power supply units by CAEN
\footnote{CAEN – Costruzioni Apparecchiature Elettroniche 
Nucleari S.p.A., Via della Vetraia, 11, 55049 Viareggio 
(Lucca), Italy.} are used. 
We have selected power supplies of the line hosted in the SY4527 
mainframe~\cite{caen-sy4527}, which offers a compact  electrical supply 
to the HV units and the possibility of fast access to the monitoring options 
via a Gigabit Ethernet interface present in the mainframe control unit. The 
control of the power supply units is also performed via the same interface.

THGEM electrodes as well as the protection wires, the drift wires and the 
lateral auxiliary electrodes are powered by A1561HDN 12-channel
modules~\cite{caen-A1561HDN} capable of voltage supply down to -6~kV. 
They are fully floating power supplies, therefore, we can use a 
very convenient on-detector ground reference, namely the detector frames themselves. 
The most relevant features of these units for our application are the low voltage ripple, 
typically 5~mV pp from 10~Hz to 100~MHz at full load, the fine voltage resolution, 
100~mV for setting and 10~mV for monitoring, and the fine current resolution,
500~pA for setting the maximum current and 50~pA for monitoring. These figures, 
which are quoted in the specifications, have been confirmed in operation.

MM pads are supplied by A7030DP 12-channel modules~\cite{caen-A7030DP}, with a maximum 
voltage of +3~kV. Also these modules are fully floating. The noise figures are $<$15~mV typical, 
20~mV maximum in the range 10-1000~Hz and $<$5~mV typical, 10~mV maximum above 1000~Hz. 
The voltage resolution is 50 mV for setting and 10 mV for monitoring. These modules offer 
less fine current resolution, which is 2~nA for monitoring. A variation of the off-set 
of the current monitoring has been observed during operation.

The power supply modules are housed in two mainframe SY4527 units, one placed near
the top detectors and one placed near the bottom ones to minimize the length of the 
high voltage cables; with this arrangement, 5~m long cables are used.

\begin{figure}
\begin{center}
\resizebox{0.5\textwidth}{!}{ \includegraphics{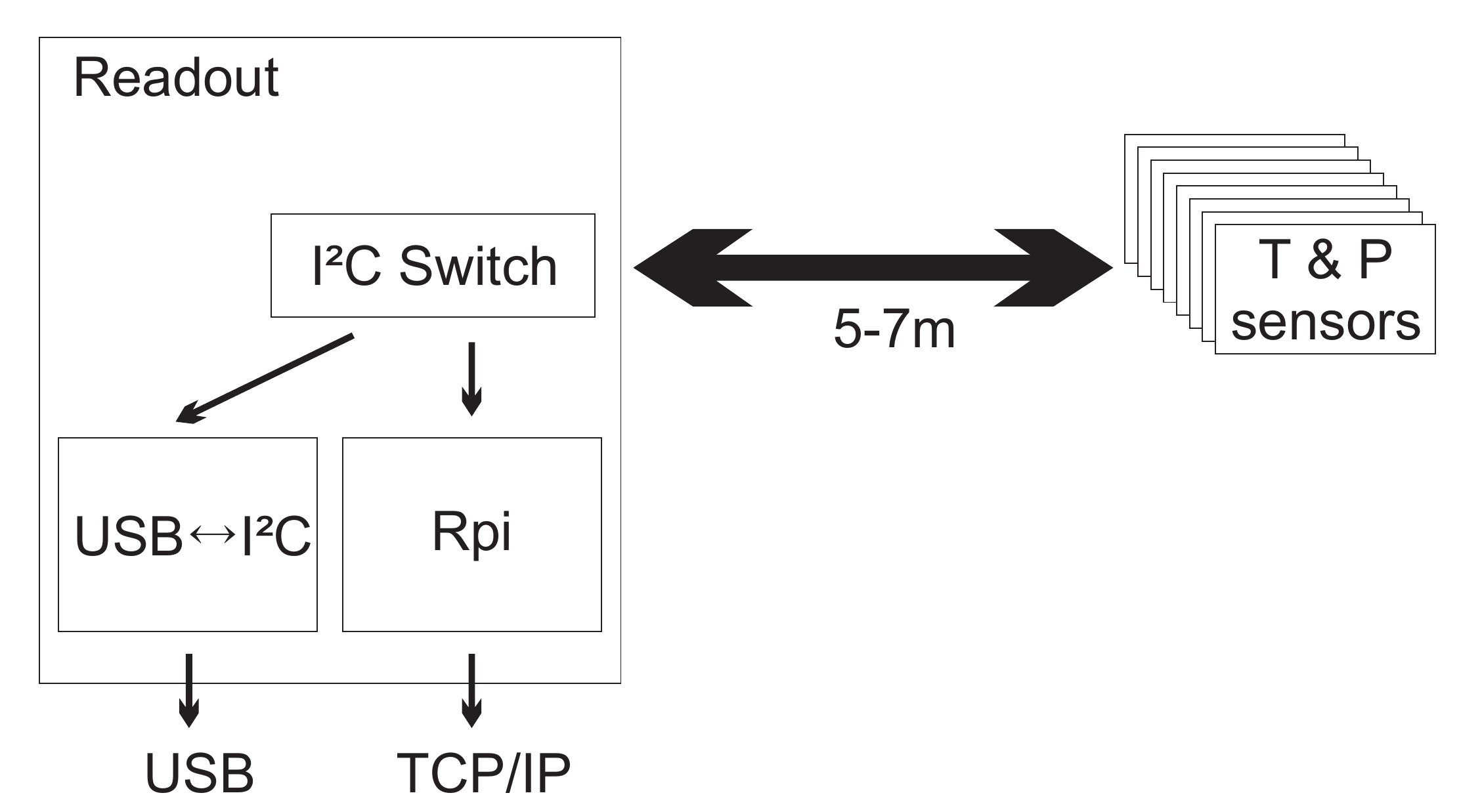} }
\caption{Block diagram showing how the data is read out from sensors.
}
\label{fig:blocks}
\end{center}
\end{figure}

\begin{figure}
\begin{center}
\resizebox{0.45\textwidth}{!}{ \includegraphics{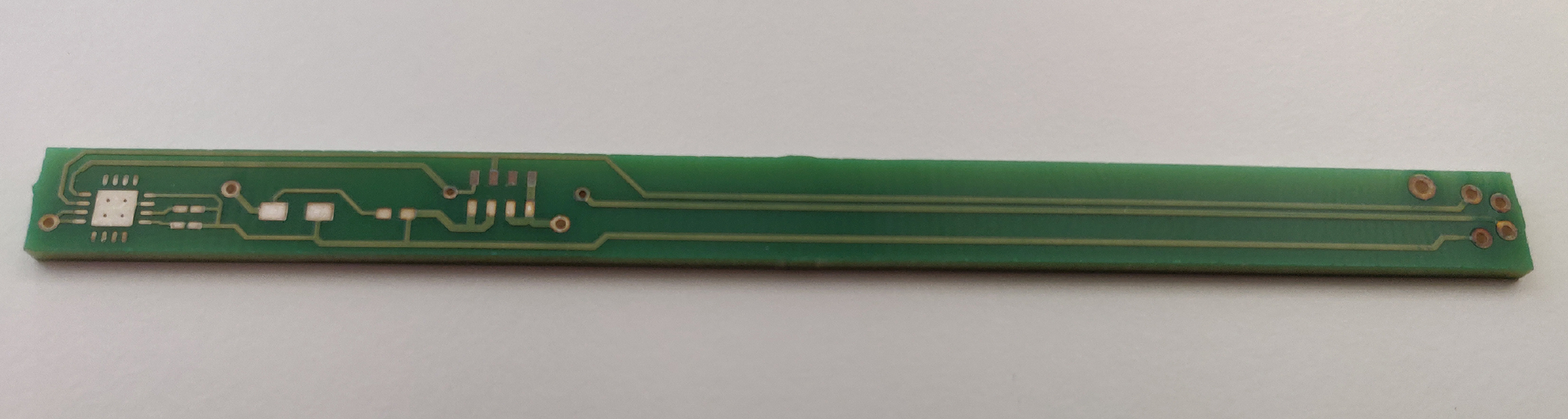} }
\caption{Picture of the PCB housing the pressure and temperature sensors.
}
\label{fig:nose}
\end{center}
\end{figure}

\subsection{Temperature and pressure measurement system}
\label{sec:p-t-system}

The main components of the system to measure temperature and pressure are the 
temperature and pressure sensors and the interfaces that make the data 
available in parallel both to a Single Board Computer (SBC) and to a Linux PC
(Fig.~\ref{fig:blocks}).

Eight couples of sensors, for temperature and pressure, are used, each mounted 
on a dedicated board, installed in direct contact with the gas flowing 
inside the chambers. A reasonably fast following of the temperature changes is
obtained by reducing as much as possible the thermal mass of the board. 
The cable connectors must be gas-tight to avoid gas contamination and leakage.
Following these requirements, sensors were mounted on the top of a very narrow 
PCB (100~mm $\times$ 7.5~mm), together with an IP67 4-pin M8 
connector to allow sensors readout (Fig.~\ref{fig:nose}) . The PCB mechanical support
allows direct mounting on the chambers' frames.

The selected sensor for temperature measurements is the ADT7420 \cite{REF1} digital 
sensor by Analog Devices, with 16-bit $\pm$ 0.25$^{\circ}$C accuracy and I$^2$C 
digital output. The choice for pressure measurements is the MS5611-01BA03 \cite{REF2} 
barometric pressure sensor with an accuracy of $\pm$2.5 hPa, TE connectivity  
and an I$^2$C digital output. The choice of these two devices was mainly due to 
the experience gained during previous developments: they fulfill the requirements 
on accuracy, stability and robustness.

Eight couples of sensors have to be read, while the sensors themselves have just one 
or two I$^2$C address bit: it is not possible to mount all the sensors on a single bus. 
Therefore, multiple I$^2$C buses have been used. They are multiplexed via a Texas 
Instruments TCA9548A 8-channels I$^2$C switch \cite{REF3} and each sensor couple is served 
by an independent I$^2$C bus. The sensor boards are connected to the readout board 
via 4-cores cables which are up to 7~m-long. The readout frequency is low. Tests and, later, 
standard operation show that no I$^2$C bus data loss was introduced by the cable length.
The TCA9548A switch output is connected both to the Raspberry Pi 3 Model B SBC~\cite{REF5}
via its GPIO and to a 
converter FT232H single channel USB to multipurpose UART/FIFO IC by FTDI \cite{REF4}. 
Therefore, it is possible to read out the sensors using the SBC, sending out data via its 
on-board WiFi, or to read out the data by a computer using 
its USB connector directly.

In the COMPASS RICH the measured temperature and pressure values are read out locally 
via the RaspberryPi. The regular readout is followed by a 
validity check for all the sensors, and if approved are sent to the HV Control 
System (Sec. \ref{sec:hv-control}).

\begin{figure}
\begin{center}
\resizebox{0.5\textwidth}{!}{ \includegraphics{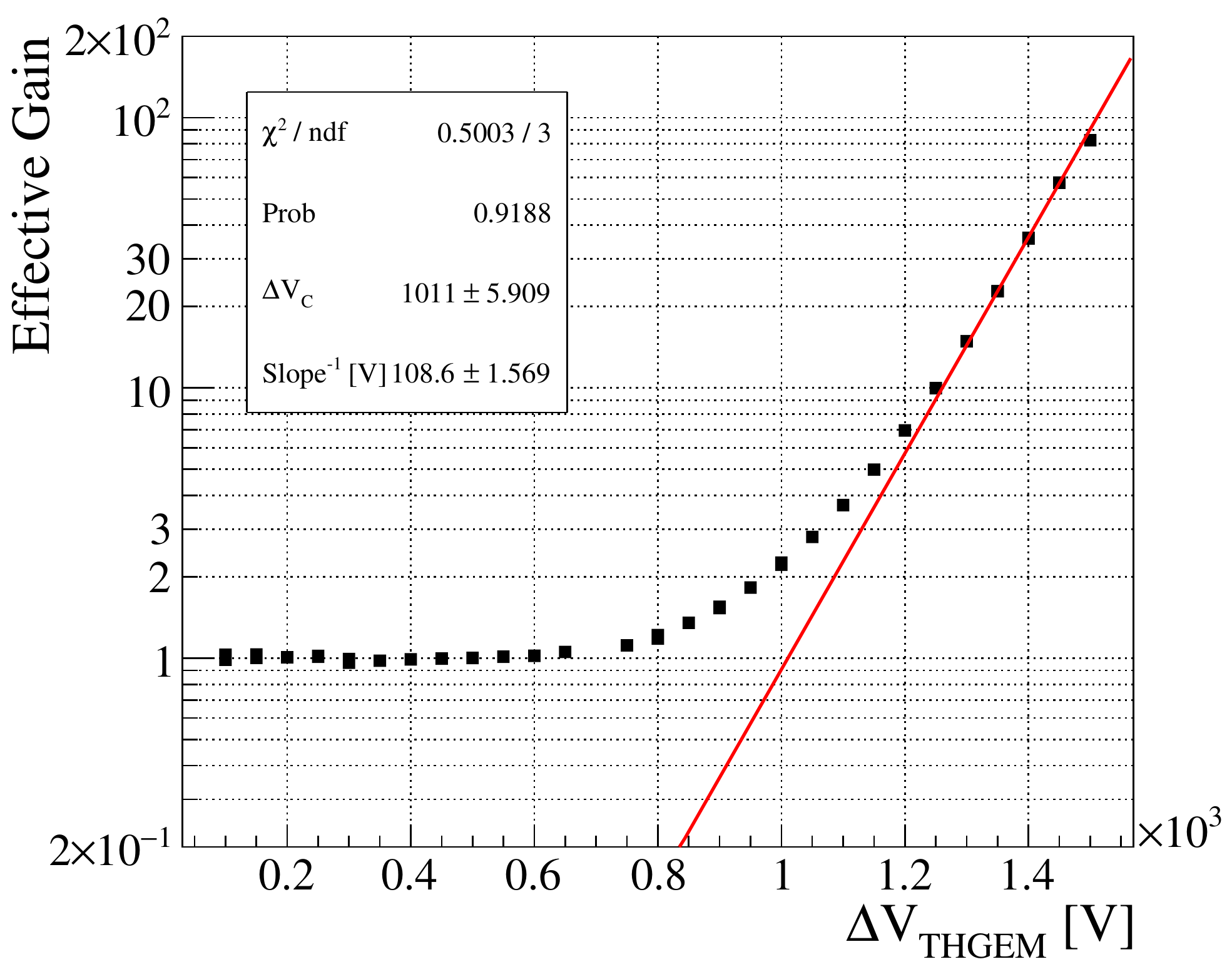} }
\caption{Gain curve of a THGEM with the geometrical parameters 
reported in Sec.~\ref{sec:architecture}
operated in Ar:CH$_4$ 50:50. The red line is a fit to 
the data using equation~\ref{eq:gain}, for the data above 1.2~kV.
}
\label{fig:THGEM_gain_calibration}
\end{center}
\end{figure}

\begin{figure}
\begin{center}
\resizebox{0.5\textwidth}{!}{ \includegraphics{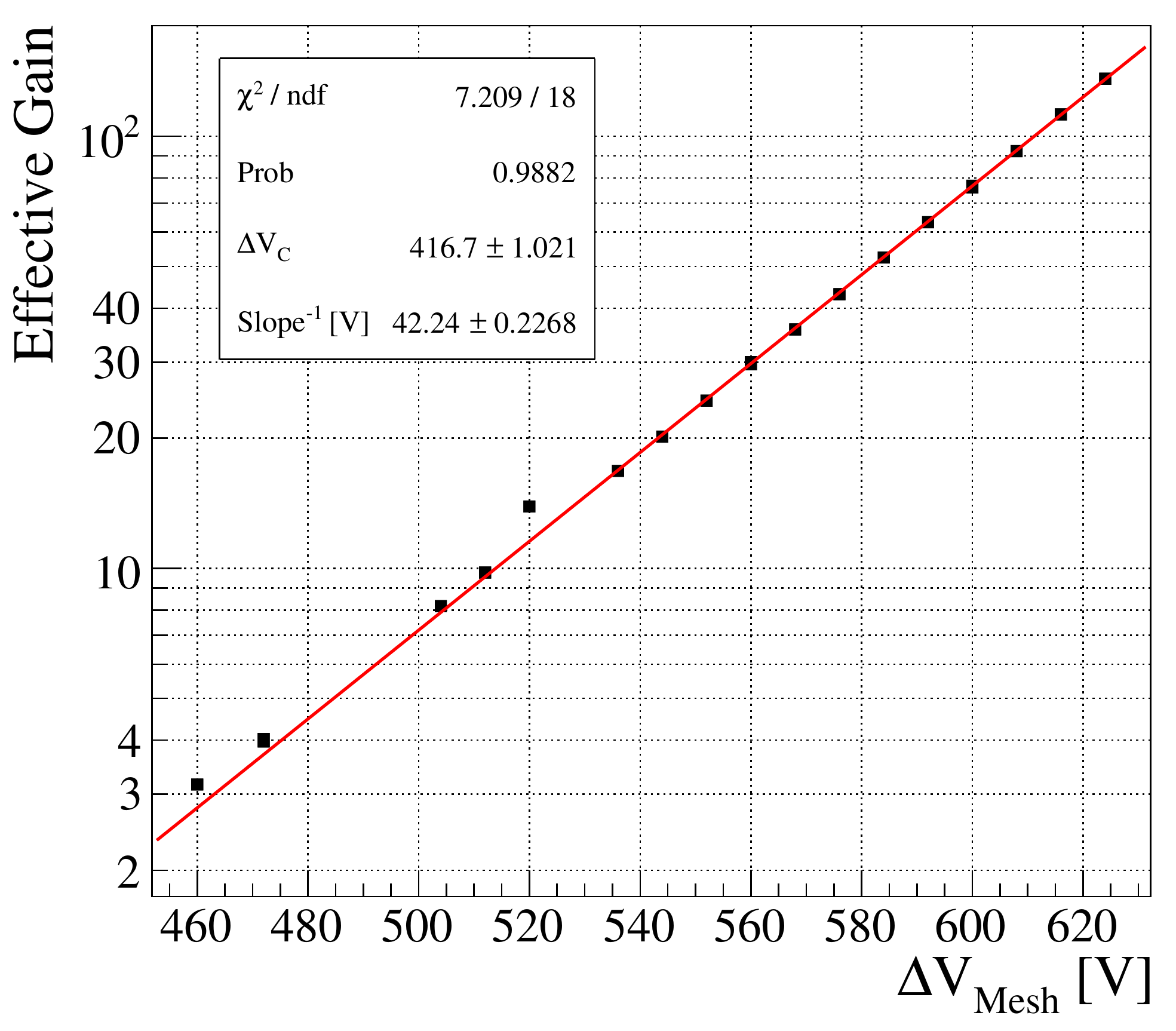} }
\caption{Gain curve of a MicroMegas with the geometrical parameters 
reported in Sec.~\ref{sec:architecture} operated in Ar:CH$_4$ 50:50. The 
red line is a fit to the data using equation~\ref{eq:gain}.
}
\label{fig:Mesh_gain_calibration}
\end{center}
\end{figure}

\begin{figure}
\begin{center}
\resizebox{0.48\textwidth}{!}{ \includegraphics{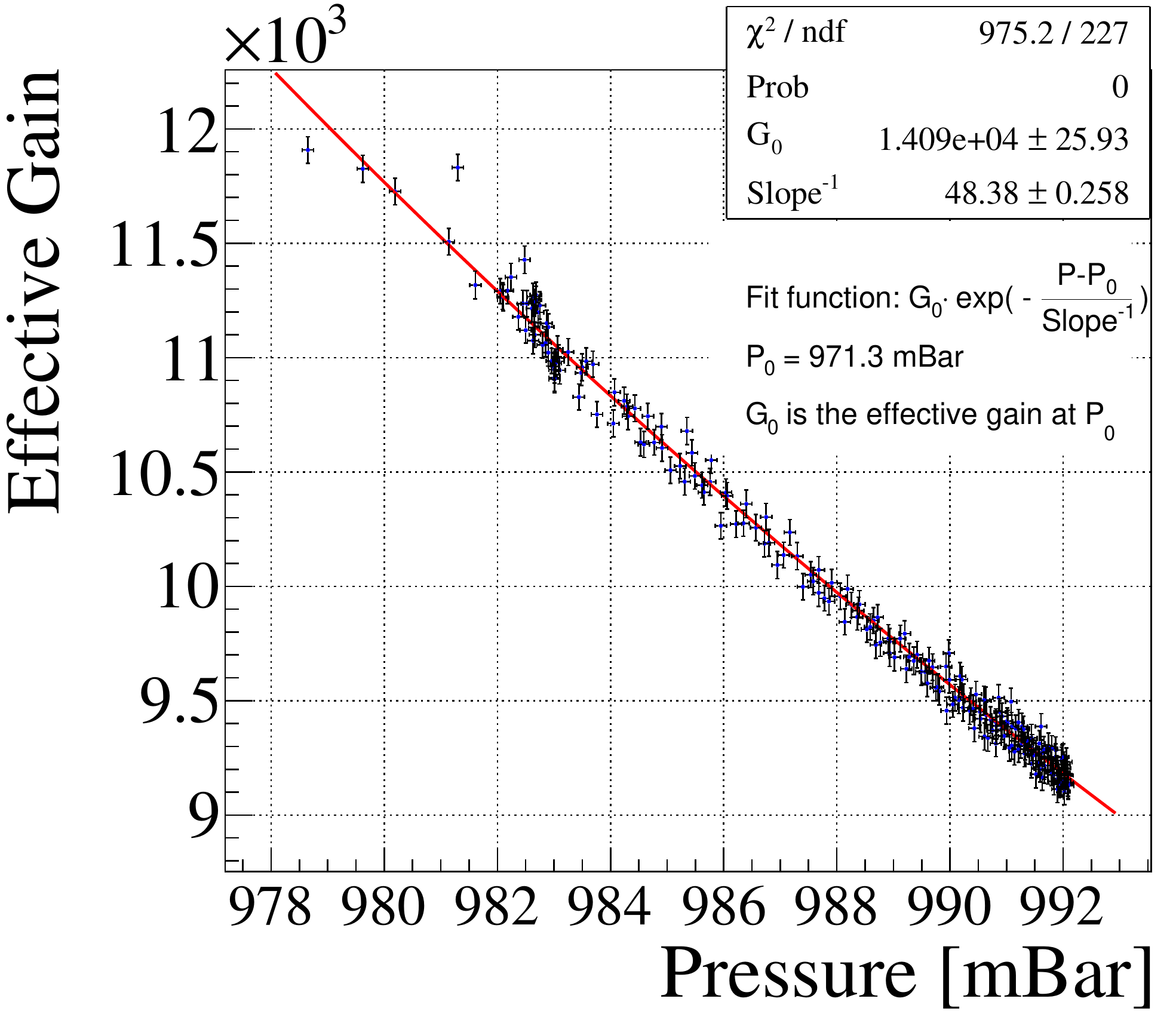} }
\caption{Gain dependence on the pressure. The working gas is Ar:CH$_4$ 50:50. 
}
\label{fig:gain_pressure_calibration}
\end{center}
\end{figure}

\begin{figure*}
\begin{center}
\resizebox{0.75\textwidth}{!}{ \includegraphics{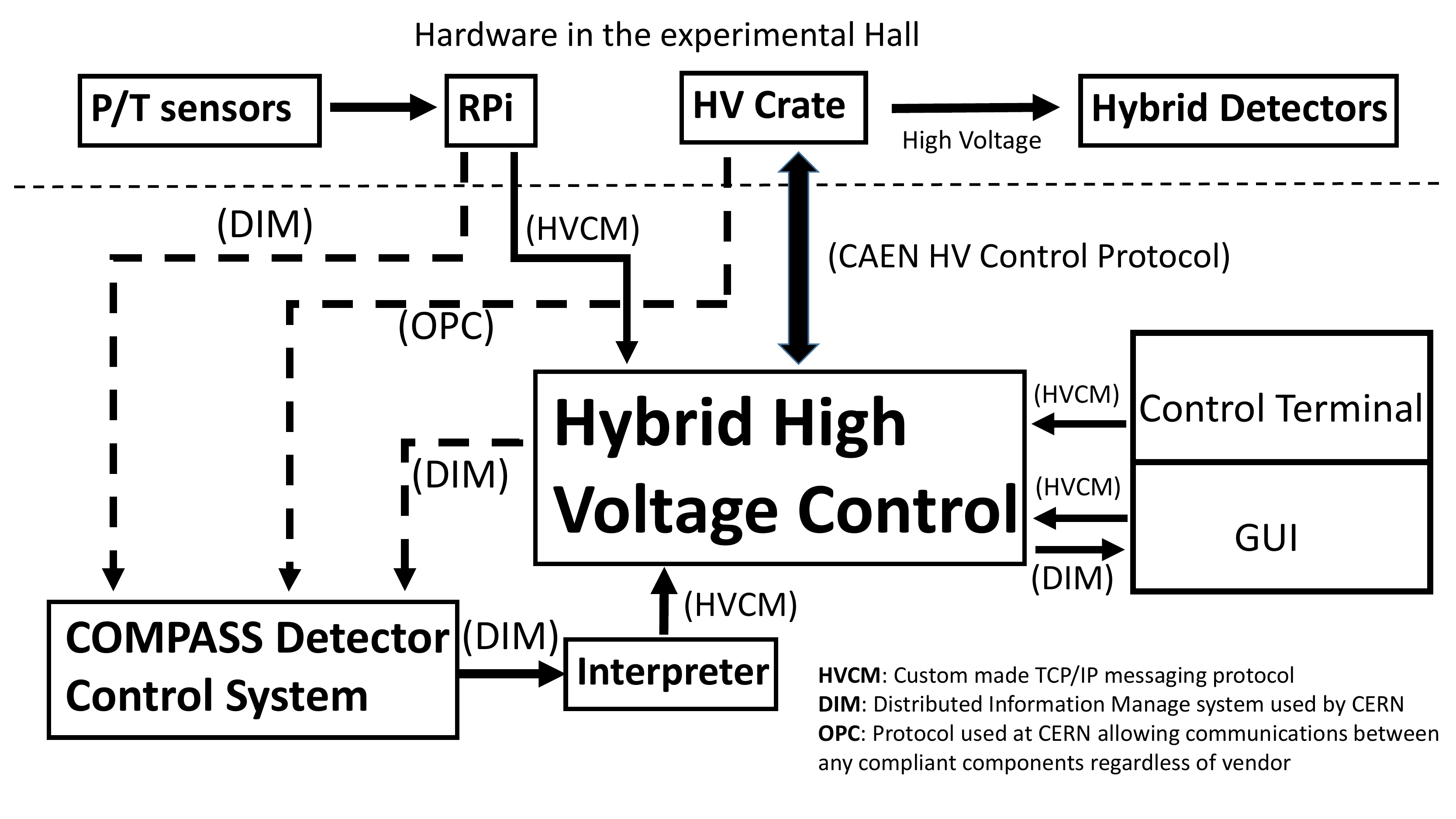} }
\caption{Block diagram of the hybrid high voltage control system, showing the components 
and their communication channels.
}
\label{fig:HV_control}
\end{center}
\end{figure*}

\section{Gain calibrations studies}
\subsection{Gain dependence}
\label{sec:gain_dependence}

In order to understand the gain performance of the hybrid detector, THGEMs and MMs have 
been extensively characterized by laboratory studies. In general, following the studies
by Townsend \cite{Sauli:117989}, the gain G has an 
exponential dependence on the bias voltage applied to a gaseous detector via the first Townsend coefficient: 
\begin{equation}
    G \sim e^{\frac{ \Delta V - \Delta V_{C} }{slope^{-1}} },
    \label{eq:gain}
\end{equation}
where $\Delta V_{C}$ is the critical voltage where the extrapolated gain is equal to 1. For 
THGEM with the geometric parameter reported in Sec.~\ref{sec:architecture} and 
using a gas mixture Ar:CH$_4$~=~50:50, the $slope^{-1} \approx 100$ 
(Fig. \ref{fig:THGEM_gain_calibration}), namely, above $\Delta V_{C}$, an 
increase of 1~V will result in a 1\% gain increase. While for MM with the 
geometric parameter reported in Sec.~\ref{sec:architecture} and using a gas 
mixture Ar:CH$_4$~=~50:50, the $slope^{-1} \approx 50$ 
(Fig.~\ref{fig:Mesh_gain_calibration}), i.e., above $\Delta V_{C}$, an increase 
of 1~V will result in a 2\% gain increase.

\subsection{Gain variation induced by environmental parameters}
\label{sec:gain_variation}

The dependence of the gain from the environmental parameters, namely temperature and pressure,
was studied for the hybrid detector as a whole. This exercise provides useful information 
only if the three multiplication stages are operated at relative gain-values near to those used 
in the operation at the experiment, because the slopes of the gain function for THGEMs 
and MMs are different.

In gaseous detectors, the gain variation due to the environmental parameters is caused by 
the variation of the gas density, which increases linearly with the pressure P and is 
inversely proportional to the absolute temperature T. 
Therefore, considering a limited range of parameter variations, the gain can be regarded as 
a function of T/P.
The function has been determined by laboratory exercises. Measurements performed at fix 
pressure and varying the temperature, provide interesting indication, but are not fully 
reproducible. In fact, the temperature of the gas in the detector is a field, which is 
only partially related to measurements such as the room temperature around the chamber, the 
temperature of the gas injected in the detector or the gas coming out from it. On the 
contrary, the gas pressure inside the detector is with very good approximation the 
same everywhere. Correspondingly, measurements performed at fixed T and varying P provide 
more reliable information.

The measurements of the gain dependence versus the gas pressure have been performed 
illuminating the detector with a $^{55}$Fe source; the pressure at the gas-outlet line of the
detector was measured with 0.1~mbar resolution; amplitude spectra have been collected.
The pressure in the chamber was increased after each gain measurement by increasing the oil 
level in the safety bubblers downstream of the gas output line. The series of measurements 
have been performed in a short time interval, so that temperature variations were negligible.
Various sets of measurements have been collected at different temperature providing consistent results.
The gain evolution versus P is illustrated by one of the sets of measurements shown in
Fig. \ref{fig:gain_pressure_calibration}: a 10~mbar P variation causes a 20\% gain 
variation. In the measured region, the dependence is approximately linear and its slope 
can be extracted from the data. The resulting voltage correction algorithm is:
\begin{equation}
    \frac{{\Delta}V}{{\Delta}V_0} = 1 + 0.55(\frac{P}{T} \frac{T_0}{P_0} -1),
\label{correction_equation}
\end{equation}
where $\Delta$V is the voltage bias applied to THGEMs or MMs and $\Delta$V$_0$, P$_0$ and T$_0$are 
fixed reference values. In particular, P$_0$ = 1000 mbar and T$_0$ = 300 K, V$_0$ is
the preset voltage for each electrode. In the hybrid detector 
architecture, the bias voltages of all the three multiplication stages have to be 
modified to obtain the compensation. Equation~\ref{correction_equation} provides 
the correction in term of a voltage correction factor to be applied to the reference 
voltage setting $\Delta$V$_0$ of each multiplication stage.

The relevance of the compensation is illustrated by two quantitative examples. A temperature 
variation of 5 K, requires a voltage compensation of about $\sim$1\% for
each multiplication stage, while it would 
cause a gain variation of $\sim$40\% if not compensated. Similarly, a pressure variation
of 10~mbar can be compensated by a $\Delta$V of $\sim$0.5\%, while it would cause a gain 
variation of $\sim$20\% if not compensated.

There are practical limitations to the effectiveness of the correction procedure.  It is 
difficult to access the gas temperature in a large detector, where the temperature is 
not homogeneous. We measure the temperature inside the detector volume near to the gas 
inlet and outlet (Sec.~\ref{sec:p-t-system}) and average the two values in order to minimize 
this effect.

Temperature and pressure values are read out at 1~Hz frequency.
Voltage correction is applied every 10 min assuming the
average values of temperature and pressure from the last 10 min.

\section{The high voltage control}
\label{sec:hv-control}

A dedicated High Voltage Control System (HVCS) software package was designed 
in order to power the COMPASS RICH Hybrids, taking into account the 
requirements as specified in Sec.~\ref{sec:requirements}. In particular, it handles 
more than hundred HV electrodes, it continuously performs the gain correction 
required by the modification of the environment parameters, checks the detector's 
electrical stability and reacts to the possible detectors misbehaviour, connects 
to the slow control system of the experiment (named Detector Control System, DCS)
and offers the experts control and log of the full detector system.

The schematic block diagram of the whole HV system is shown in Fig.~\ref{fig:HV_control}, 
where the communication protocols are highlighted.

The HV power supplies housed in the dedicated HV SY427 crates (Sec.~\ref{sec:power-supply}) 
are controlled by the Hybrid HV Control (HVC) software package and using
the Ethernet port of the crate. HVC addresses the HV channels in the CAEN HV crate using the 
official CAEN Wrapper libraries. Meanwhile, the monitoring by DCS via the direct access 
to the crate-resident OPC server is uninterrupted. HVC is written 
in C/C++ language in order 
to have low-power, high-performance and high-compatibility.
\par
The communication between the system parts, the DCS, and the users are via messages, 
using the DIM protocol \cite{DIM} and a simple TCP/IP socket protocol 
named ``hvcm", developed at INFN Trieste.
Messages are used to send P and T information from 
the RaspberryPi, placed in the experimental hall. At the same time, messages are used to send 
information from and to the COMPASS DCS. The actuation by the users and experts
is also via messages.

A frequent check and log of voltages and currents are performed at 1~Hz, in order to 
realize detailed monitoring of the system, in particular of dark currents, and to 
promptly detect sparks and electrical instabilities. Later, data can be analyzed in 
details (Sec.~\ref{sec:analysis}), while spark-counting is performed on-line.
Sparks cause spikes in the current; in case a detector sector sparks at a rate higher 
than accepted  (maximum of 5 sparks per hour), the HVC system automatically decreases 
its voltage, sends the corresponding information to the DCS, and issues information 
e-mails to the on-call operator and the experts.

Handling a swarm of channels, several correlated among them, requires a simple and 
functional scheme. The detector has 
a well defined ideal voltage setting, assumed as default. The voltages on the real 
individual sectors and channels are scaled compared to this one. Three scales (multiplication factor) are 
used, named : \textit{SacleSet}, \textit{ScalePt}, and \textit{ScaleOwn}.

The \textit{ScaleSet} parameter applies as a general multiplier to all the electrodes 
of a sector, and is used to access the normal/safe/idle voltage settings, without 
turning the HV off.

The scaling applied for the required voltage compensation due to a change in environmental parameters
are applied via the \textit{ScalePt} parameter. This multiplier is obviously applied only to the 
amplification stages (THGEM1, THGEM2, MM).

THGEMs and MMs are not identical due to production limitations and can also be non-uniform. 
Fine adjustment of the applied voltages of the individual stages helps to achieve a uniform 
gain throughout the active surface. In case, any electrically feeble segment is present at
a particular stage, the sparking probability of that stage can be reduced by decreasing 
the voltage of that segment. However, the lower voltage reduces the gain of that particular 
segment. The gain compensation can be made by obtaining a higher gain value at the 
corresponding segment of a different stage. The fine adjustment of the applied voltage 
is therefore used to change the gain sharing of the stages for each sector. This 
fine-tuning of each of the amplification stages is obtained by applying 
the \textit{ScaleOwn} parameters.  


An excellent gain uniformity has been reached over the segments of the detectors by the fine adjustment of the single amplification stages (Sec.~\ref{subsec:gain_uniformity}).

\begin{figure}
\begin{center}
\resizebox{0.55\textwidth}{!}{ \includegraphics{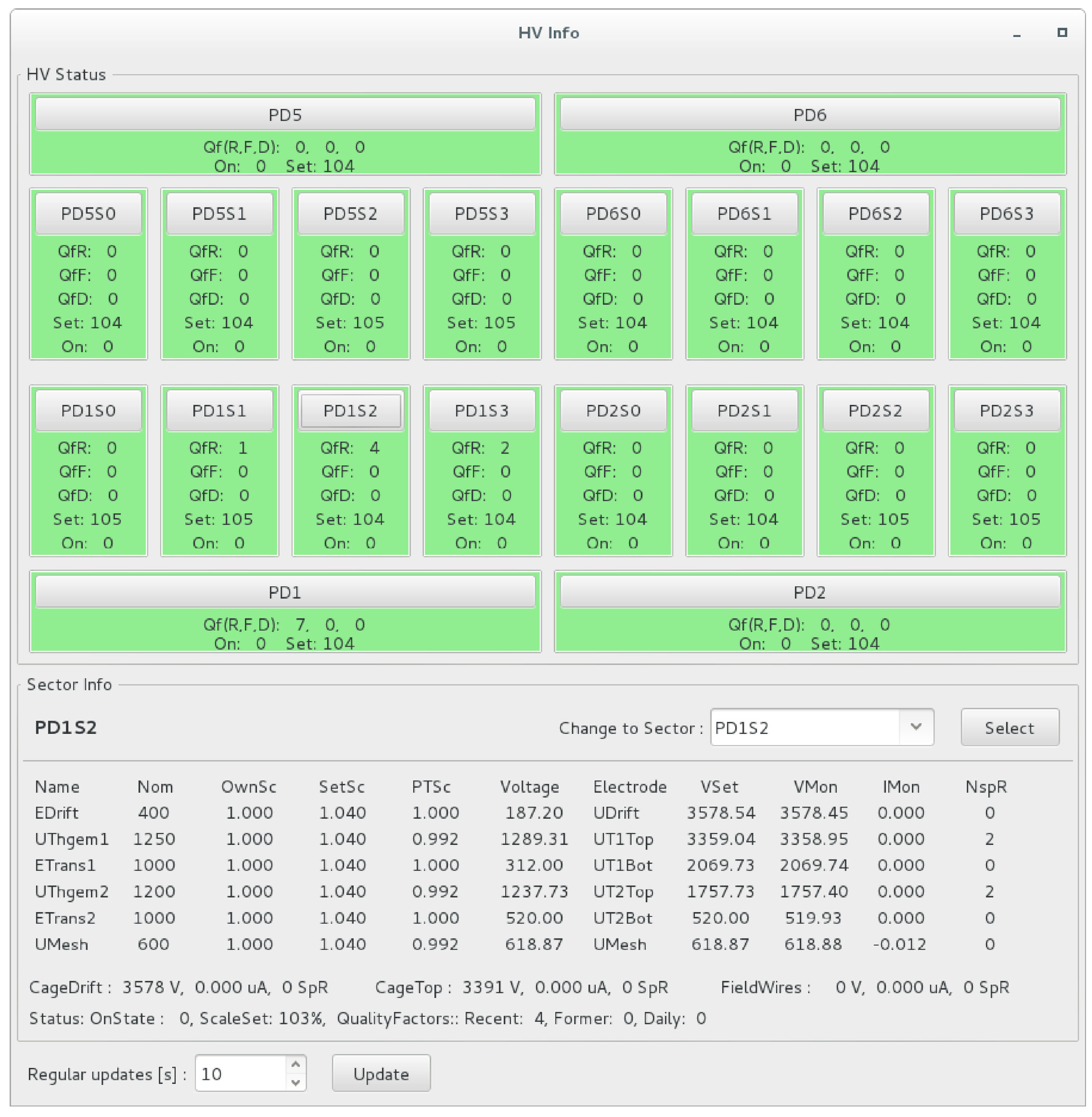} }
\caption{Screen shot of the main frame of the 
graphical interface, providing information on all 
the detector sectors, the 
spark counts, and details of the selected PD1S2 sector.
}
\label{fig:gui}
\end{center}
\end{figure}

There is no need of interaction from the users during ordinary data taking because everything 
is automated. On the contrary, during the commissioning phase, a thorough monitoring of the 
operational parameters and a platform for interactions is needed. A graphical user interface 
has been written: it reads the parameters of the sectors, including spark counting and actual 
scales, and shows them in user friendly colour-coded frames that also 
acts as an interface to set changes of the scales or of the nominal parameters (Fig. ~\ref{fig:gui}). The graphical 
user interface is written in wxWidgeds.

\section{Analysis of the high-voltage data}
\label{sec:analysis}

The data logged by HVCS have been analyzed. They include voltages and currents 
of the different electrodes, as well as temperature and pressure information at the four detectors.
The results have been related to the information concerning
the detector gain extracted from the COMPASS 
spectrometer data. In the following, we report the analysis of the data collected during the 
2017 COMPASS run, about 80 days long.

\subsection{Gain estimation from COMPASS spectrometer data}
\label{subsec:gain}

The behaviour of the detector gain is extracted from the COMPASS spectrometer data. The 
gain estimation has to rely on a limited-size data sample in order to make possible 
the assessment of the gain evolution versus time. It also has to be simple and based 
on RICH data only in order to provide on-line monitoring of the gain evolution.

\begin{figure}
\begin{center}
\resizebox{0.48\textwidth}{!}{ \includegraphics{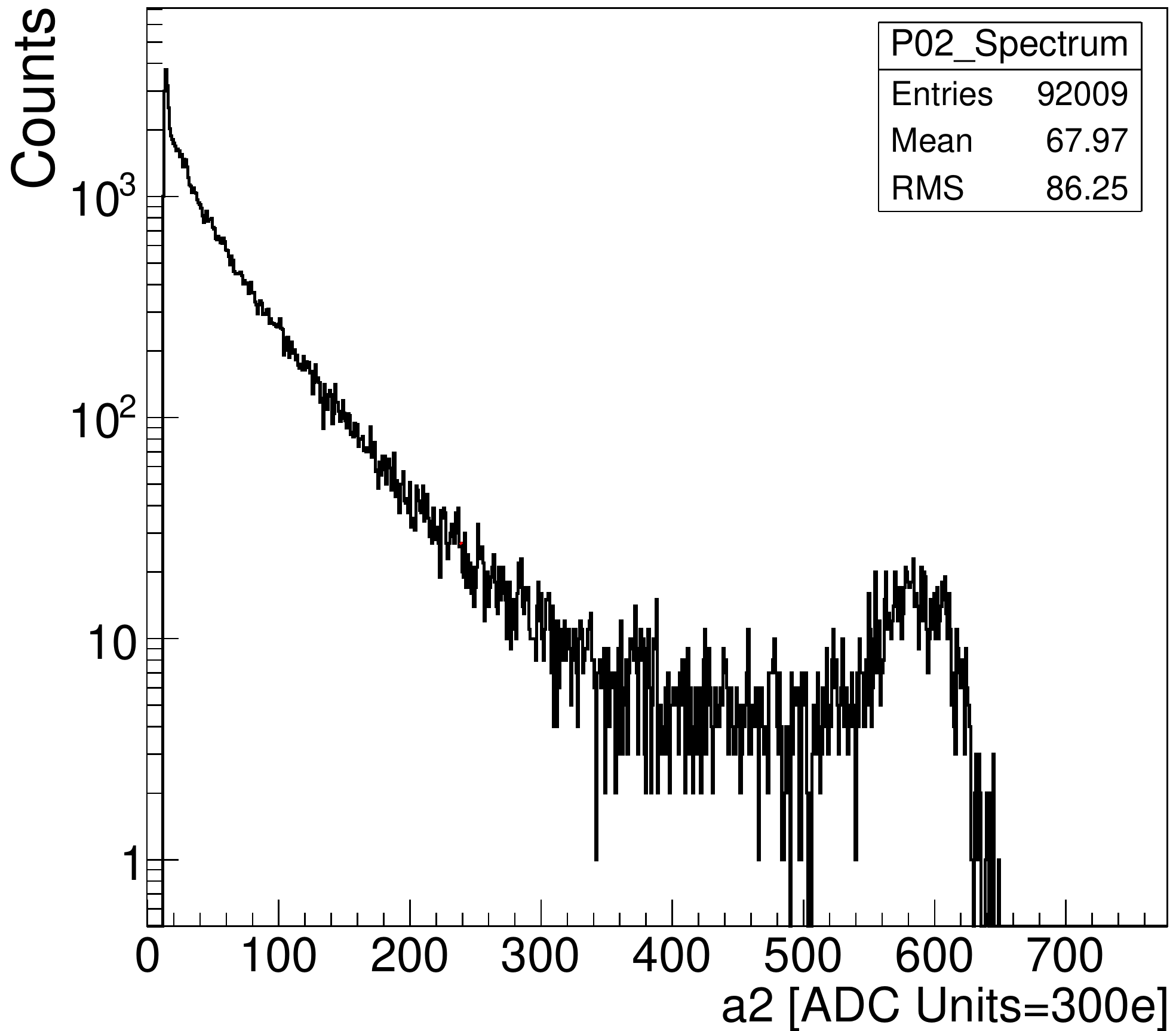} }
\caption{Example of amplitude spectrum obtained applying the data 
selection explained in the text. The amplitude units are ADC channels, 
where one channel corresponds to 300 electrons.
}
\label{fig:amplitude_spectrum}
\end{center}
\end{figure}

Sets of one-million events, typically collected in about 5~min, are used. The amplitude 
spectra of the hits collected in the photon detectors are produced, where the hits are 
selected according to the criteria reported in the following. The hit readout is 
performed via the front-end chip APV25~\cite{apv25}; individual channel pedestal 
subtraction and thresholding, as well as the subtraction of the common-mode noise, 
are applied. For hits above the threshold, three amplitudes are measured, 
namely a$_2$ in-time with the trigger and a$_1$ and a$_0$, respectively, 150~ns and 
300~ns before the trigger time. When the measured a$_2$ amplitude is generated by a 
particle in-time with the event, it is expected that a$_1$ corresponds to a measurement 
related to the signal rising, while a$_0$ should represent a measurement of the noise 
level. Therefore, the following condition is imposed to select a hit: 
0.2~$<\frac{a_1}{a_2}<$~0.8 and $a_0 < 3000$ electrons.
A typical amplitude spectrum is shown in Fig.~\ref{fig:amplitude_spectrum}.

Three regions can be identified in the spectrum: at low amplitudes an important 
contribution by the residual noise can be observed, at intermediate amplitudes the 
distribution is dominated by the single Cherenkov photon signals, exponentially 
distributed, while at larger amplitude the distribution is contributed by the 
ionization signals due to particles crossing the photon detectors; this last 
portion of the distribution is also distorted due to the saturation of the 
dynamic range of the APV25 chip. The intermediate region is used for the estimation, namely the one 
dominated by the single photon signals, where the amplitude follows with good 
approximation an exponential distribution.

A gain estimator is used 
to extract the detector gain. 
In order to be robust, three algorithms are used. They are all based on the assumption 
that the distribution is merely due to single photon signals, hence, an exponential distribution. 
The gain is the reversed of the exponential function slope.
The three algorithms are:
\begin{enumerate}
    \item 
    the gain estimator G$_m$ is calculated from the intermediate region of the distribution in a 
    fixed range (50-200 ADC channels);  
    \item
    given a distribution range, the slope can be algebraically calculated; 
    this calculation is repeated for 30 ranges, each of them 150 channel wide, 
    with range starting  channel from 40 till 69 and, correspondingly, 
    ending channel from 190 till 219; each calculation provides a value G$_c^i$, 
    with i~=~1,~2,~...,30; the estimator G$_c$ is the mean value of the 
    30 G$_c^i$-values;
    \item
    given a distribution range, the slope can be extracted from a two parameter 
    best-fit procedure; this calculation is repeated for 30 ranges, the same used to compute G$_c$; each 
    calculation provides a value G$_f^i$, with i~=~1,~2,~...,30; the estimator 
    G$_f$ is the mean value of the 30 G$_f^i$-values.
\end{enumerate}

\begin{figure}
\begin{center}
\resizebox{0.45\textwidth}{!}{ \includegraphics{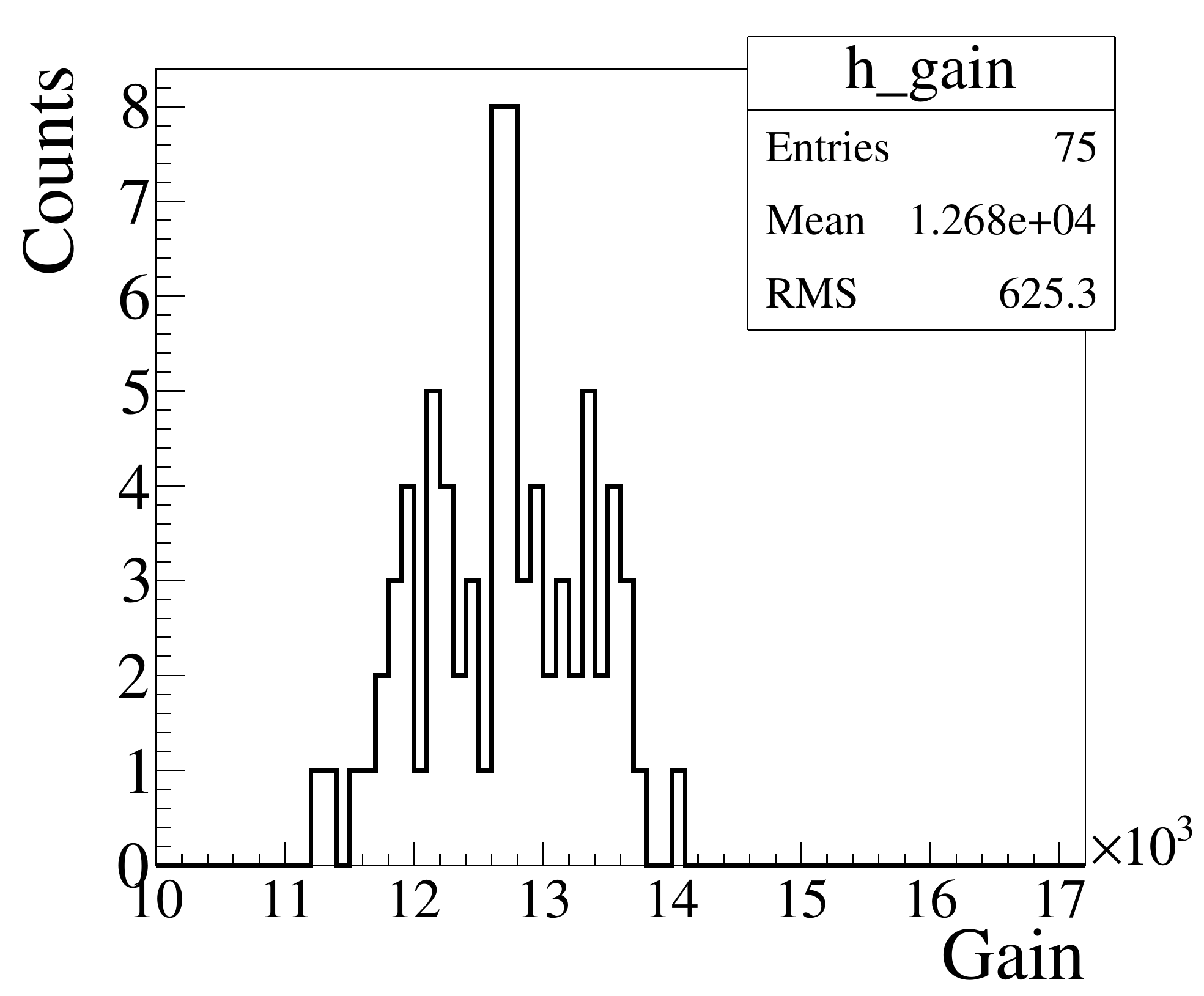} }
\caption{Estimated gain using the gain estimator G for 75 independent samples of 
data collected in conditions that ensure approximately constant gain.}
\label{fig:G_robustness}
\end{center}
\end{figure}

The final gain estimator G=G$_f$, while G$_m$ and G$_c$ are used to validate 
the result: G$_f$ is regarded as a valid gain estimation only if the differences 
between G$_f$ and G$_m$ and between G$_f$ and G$_c$ do not exceed 20\%.
Its robustness has been assessed as follows: a set of consecutive COMPASS data, 
around 20 million in total, collected in approximately 1~h,
is selected taking care 
that temperature and pressure variations are negligible and that the photon 
detectors are stable. It is assumed that the photon detector gain is constant 
for these data. The data are subdivided into 75 independent subsets of around 250k events
and the G-value is extracted for each of them. The results are shown in 
Fig.~\ref{fig:G_robustness}: the standard deviation of the distribution is 
less than 5\%, proving the robustness of the estimator.

\subsection{On-line sparks detection and monitoring}
\label{subsec:sparks}

In the hybrid MPGDs installed in RICH-1, sparks result in short current spikes, 
thanks to the resistive protection applied in the supply schemes of both THGEMs 
and MMs (Sec.~\ref{sec:architecture}).

The power supply parameters reading at 1~Hz makes it possible to know the 
duration of the spikes and the corresponding restoration time.
MM spikes have a duration of $\sim$1-2~s, while THGEM spikes 
last $\sim$10~s.  
The order of magnitude of these recovery times corresponds to the time required
for loading the equivalent RC systems.
These short duration combined with the very low spike
rate of the order of 1/h per detector (Sec.~\ref{subsec:gain-stability}) and the detector segmentation,
which limits the area affected by a spark,
makes the dead-time caused by 
detector sparks totally negligible.

\begin{figure}
\begin{center}
\resizebox{0.5\textwidth}{!}{ \includegraphics{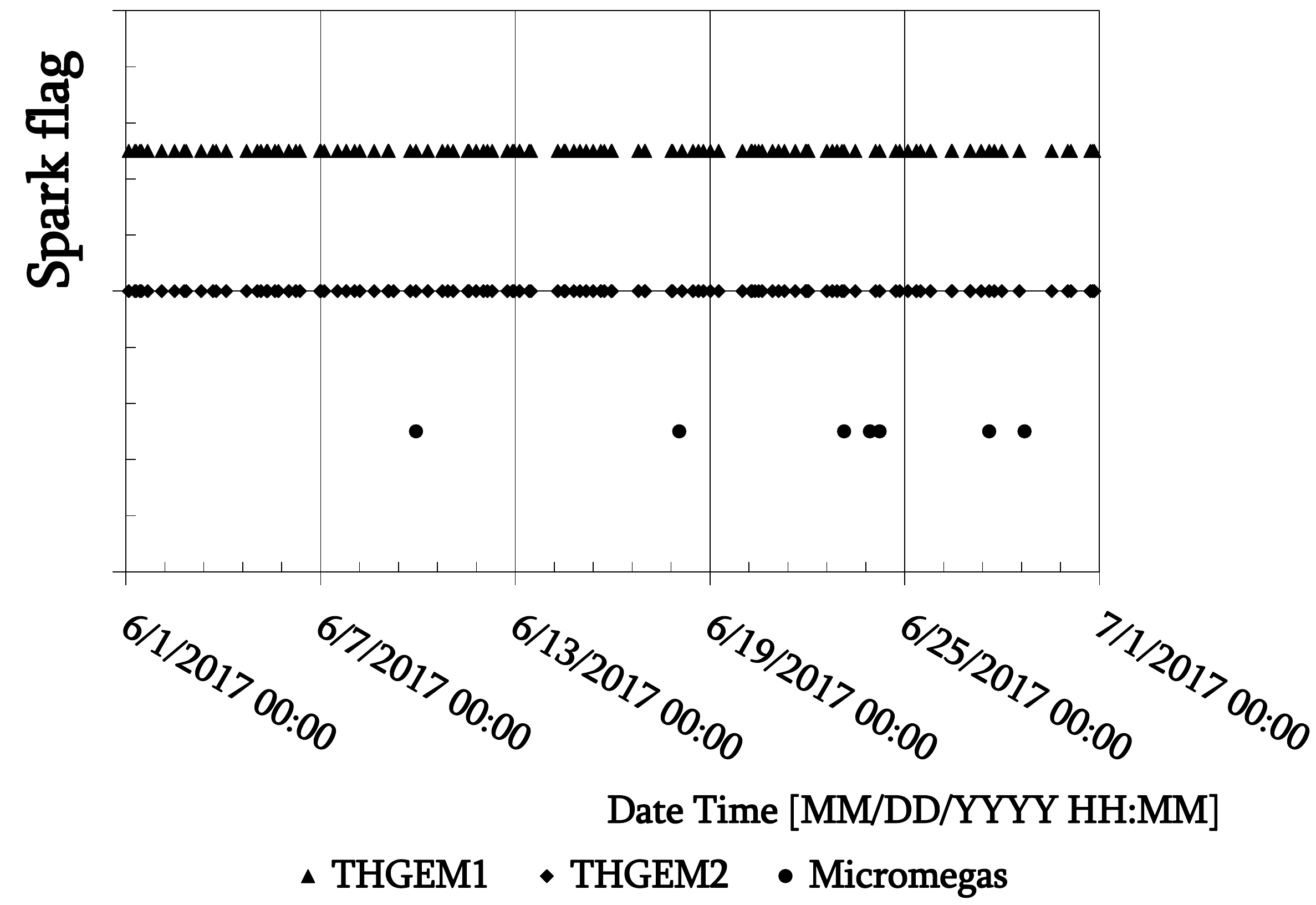} }
\caption{Correlation of the current sparks among three multiplication stages.
A marker in the top (middle) row illustrates a current spark in the THGEM1 (THGEM2).
A marker in the bottom row illustrates a current spark in the Micromegas. All the
sparks are aligned as a function of time. Data show that sparks in the two THGEMs
layers are 100\% correlated, while a spark in MM is accompanied by a spark in THGEMs
in 70\% of the cases.}
\label{fig:spark_cor}
\end{center}
\end{figure}

The sparks can be correlated among the different 
multiplication stages of a detector sector (Fig. \ref{fig:spark_cor}). 
Sparks in the two THGEM layers 
of a sector are 100\% correlated. This is expected taking 
into account that 
a spark results in a temporary short between the two THGEM electrodes,
with a drastic local variation of the electric field that affects also the 
other THGEM layer. 
The spark rate in MM is low compared to THGEMs, proving that the implemented
resistive MM scheme results in an intrinsically 
robust multiplication stage.
MM sparks are accompanied by sparks in THGEMs in 70\% of the cases.

\subsection{Gain uniformity}
\label{subsec:gain_uniformity}

\begin{figure}
\begin{center}
\resizebox{0.5\textwidth}{!}{ \includegraphics{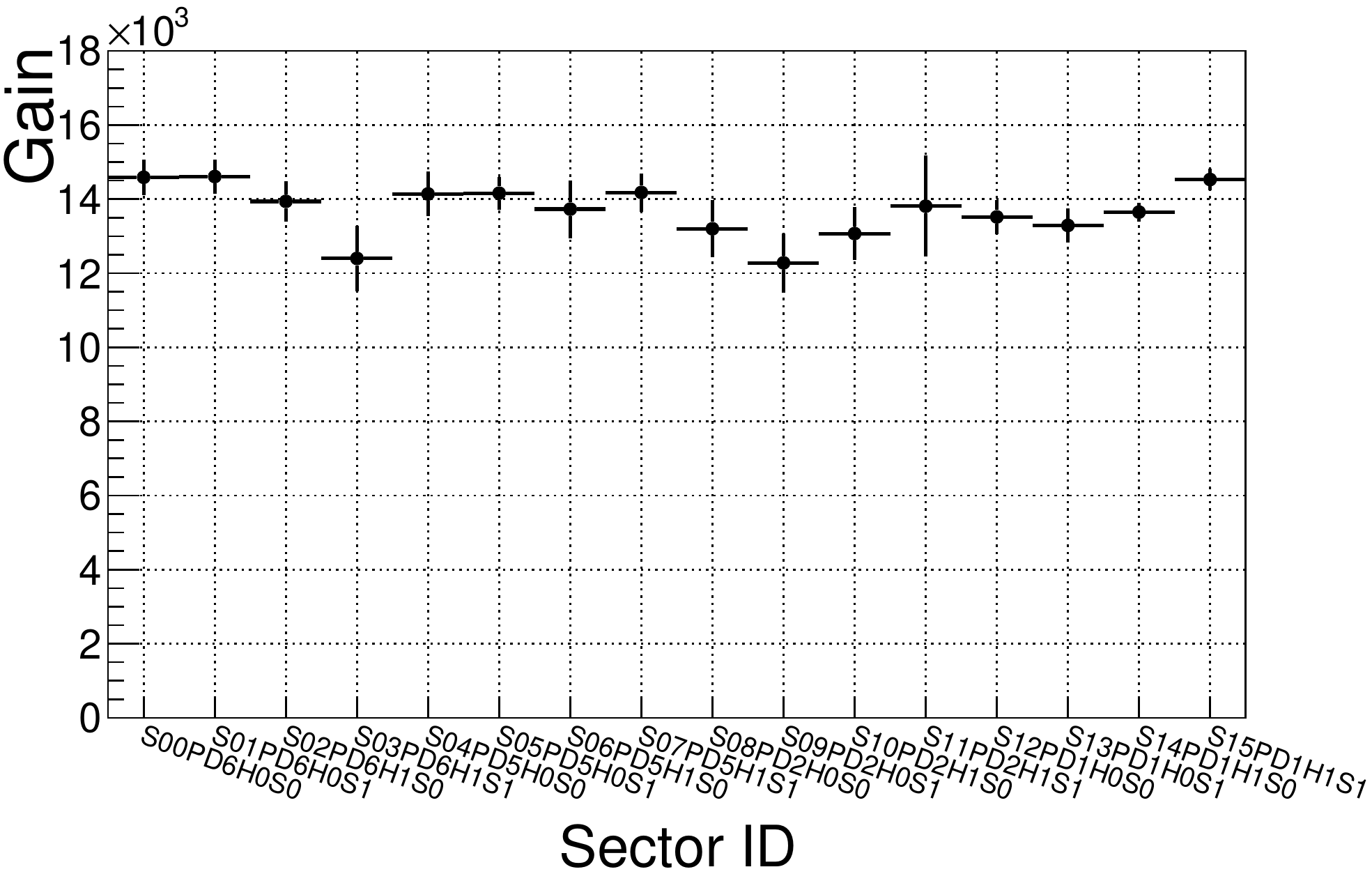} }
\caption{Gain in the sectors of the hybrid detectors. Every four points belong
to one hybrid detector, corresponding to four sectors for each detector.
}
\label{fig:gain_uniformity}
\end{center}
\end{figure}

The remarkable gain uniformity obtained with the fine tuning options 
described in Sec.~\ref{sec:hv-control} is shown in 
Fig.~\ref{fig:gain_uniformity}, where the gain is estimated  
with the algorithm described in Sec.~\ref{subsec:gain}.

\subsection{Long-term gain stability}
\label{subsec:gain-stability}

\begin{figure*}
\begin{center}
\resizebox{0.85\textwidth}{!}{ \includegraphics{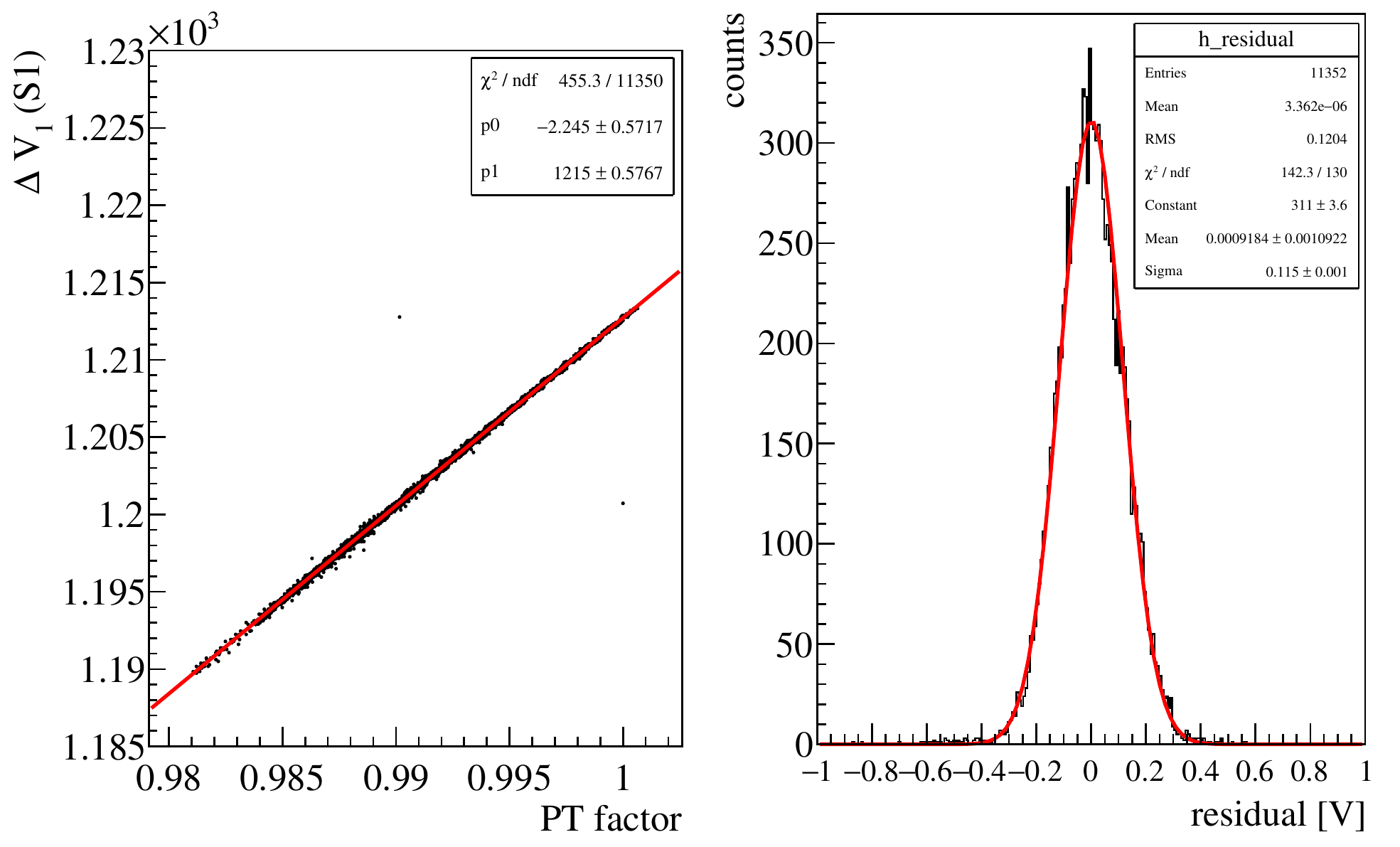} }
\caption{(Left)$\Delta V$ (unit V) of sector 1 of the first THGEM in one particular detector as
a function of correction factor. The red line is a linear fit to the data points. (right) The
residual of the data points and 
the linear fit in the left plot.
}
\label{fig:correction_PT}
\end{center}
\end{figure*}

The correct application of the voltage correction is demonstrated by the example in Fig.~\ref{fig:correction_PT}.
In Fig.~\ref{fig:correction_PT}, left,
where the voltage variation $\Delta V$ is drawn versus the voltage correction factor for the 
first THGEM of sector 1 in the photon detector 6. The points of the graph are very well aligned around 
the fitted straight line. This indicates that the linear correction due to temperature and pressure 
fluctuations is implemented correctly. The residual between the data points and the linear fit is 
shown Fig. ~\ref{fig:correction_PT}, right. The width of the Gaussian fit indicates that the 
correction resolution is around 0.1~V, matching the voltage resolution setting of the power supply
(Sec.~\ref{sec:power-supply}).

\begin{figure*}
\begin{center}
\resizebox{0.75\textwidth}{!}{ \includegraphics{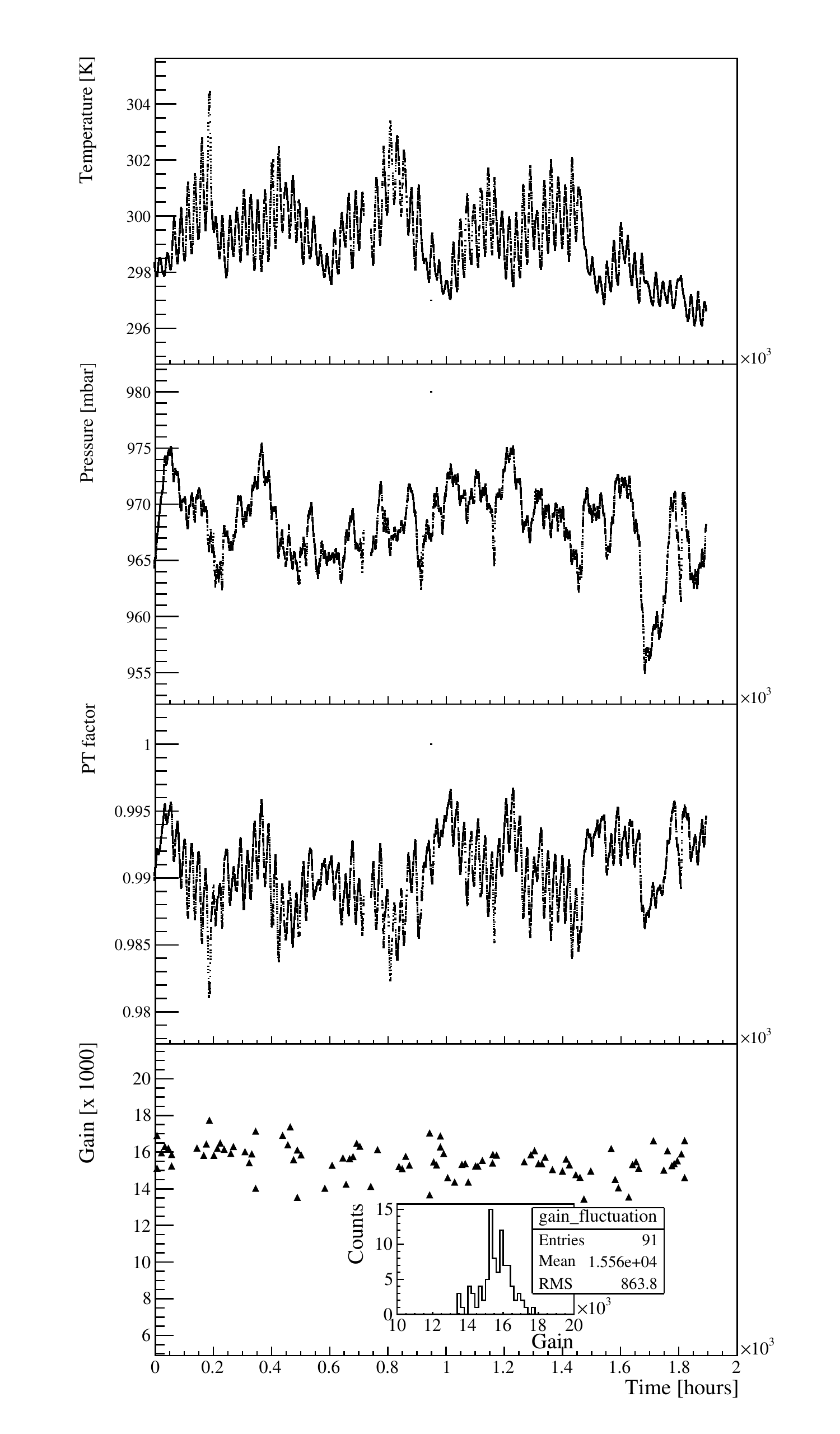} }
\caption{Temperature (first plot from the top), 
pressure (second plot from the top), the corresponding
correction factor (third plot from the top) 
and the gain fluctuations (bottom) are shown as a function of time for a continuous running of
nearly 2000 hours. 
}
\label{fig:P_T_time}
\end{center}
\end{figure*}

Fig.~\ref{fig:P_T_time} shows the temperature (first plot from the top) and 
pressure (second plot from the top) as a function of time for nearly 2000 hours 
of continuous running. The plotted values are the averaged values of the past 10~min in 
order to apply voltage corrections. The corresponding voltage correction factor 
provided by equation~\ref{correction_equation} is also shown (third plot from the top). 
The effectiveness of the correction is tested by determining the detector gain  from the 
COMPASS data using the estimator G discussed in Sec.~\ref{subsec:gain}. Two sets of data
per day are used to monitor the gain stability, one collected in the early morning and 
the other one in the late afternoon, in order to maximize the effect of the daily thermal excursion.
As an example, the gain of a sector (PD6 sector 1) is shown in Fig.~\ref{fig:P_T_time}, bottom.  
The standard deviation of the estimated gain-values is around 6\%, which includes the
intrinsic fluctuations of the gain estimator. It indicates very good gain stability.

\begin{figure}
\begin{center}
\resizebox{0.49\textwidth}{!}{ \includegraphics{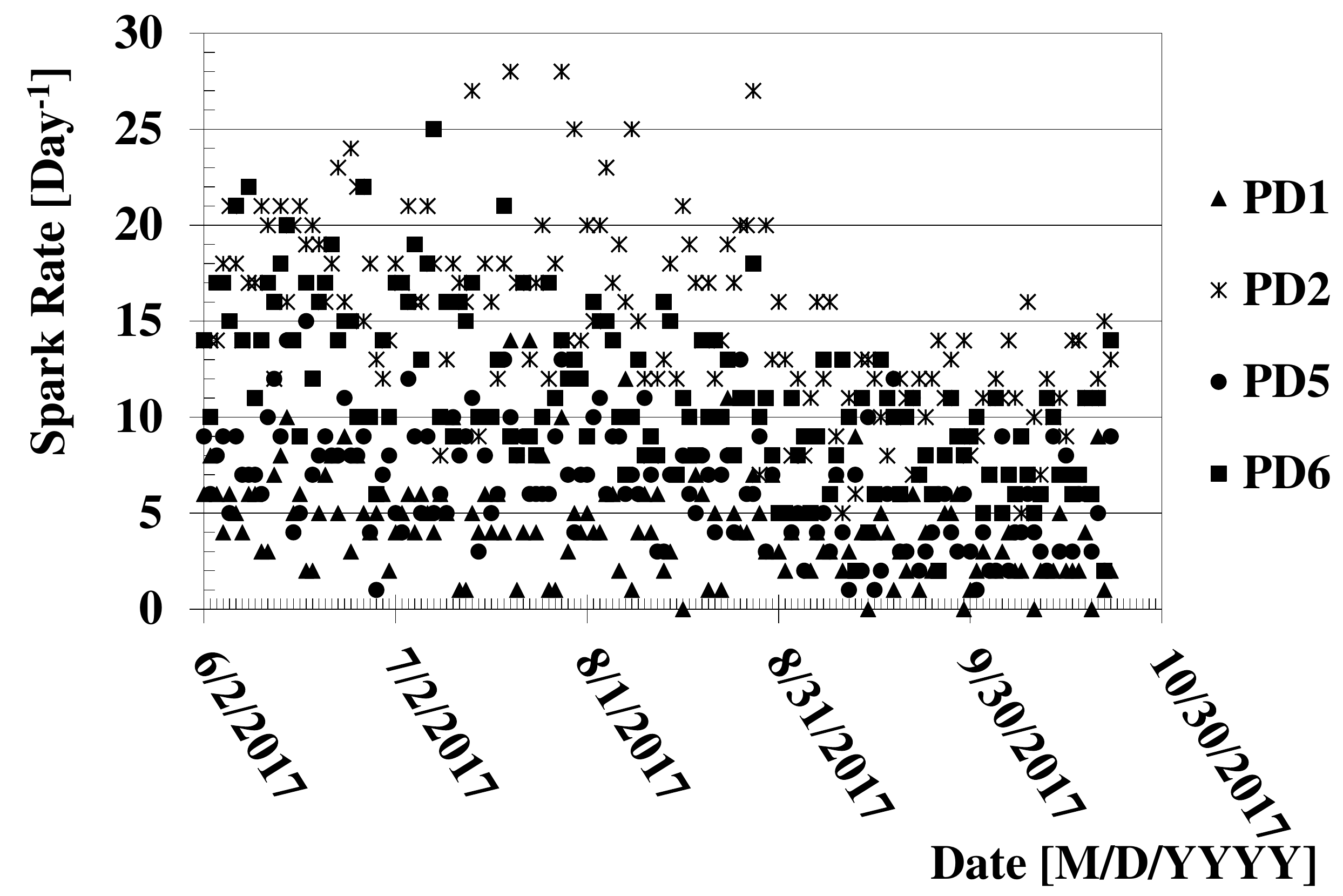} }
\caption{Spark rate counting (in unit of per day) for all four hybrid detectors during physics data 
taking of year 2017. 
}
\label{fig:sparks_rate}
\end{center}
\end{figure}

Another evidence of the stability of the whole detector system is the sparks rate in 
the detector during the physics data taking. Fig.~\ref{fig:sparks_rate} shows
the sparks rate for all the sectors of all the four new hybrid detectors from May
to October of 2017. Typically, it does not exceed 5/h in total for all four
detectors. It could be verified that there is no remarkable dependence on the beam
intensity. Sparks correlated in time have been observed for hybrid detectors, also for 
those sparks happened in sectors which are not contiguous to each other. 
These observations suggest that a major source of the observed
sparks are cosmic rays that occasionally produce relevant ionization in these detectors.

\section{Summary}
A dedicated HV system with pressure and temperature compensation has been implemented for the 
MPGD-based photon detectors of COMPASS RICH-1. Its components, namely the commercial power supply, 
the system for the monitoring of the environmental parameters and the software control system have 
been described in detail. The overall system fulfills the requirements. The system handles more 
than hundred HV electrodes, several correlated among them with a fully automatized approach.
Voltage corrections able to preserve gain stability in spite of the variation of the 
environmental parameters are continuously applied and 
the residual gain variation is at 6\% level 
over months of operation. This system is a central element contributing to the successful 
use of hybrid MPGD photon detectors in COMPASS RICH-1.

\section*{Acknowledgements}
The authors are grateful to the colleagues of the COMPASS
Collaboration for continuous support and encouragement.\\
This work is partially supported by the 
H2020 project AIDA-2020, GA no.~654168.\\

\section*{References}


\end{document}